\documentclass{aa}  

\usepackage{graphicx}
\usepackage{txfonts}
\usepackage{lipsum}
\usepackage{subcaption}
\usepackage{lscape}
\usepackage{placeins}
\usepackage[colorlinks=True,citecolor=blue]{hyperref}
\usepackage{bm,siunitx,multirow}

\begin{document}
    \title{SIRIUS Project: Dynamical Evolution of Primordial Binaries during Star Cluster Formation}
    \subtitle{}

    \author{
        Naoto Harada\inst{1}\fnmsep\thanks{Corresponding author: hrdnot-3997@g.ecc.u-tokyo.ac.jp}
        \and Michiko S. Fujii\inst{1}
        \and Takayuki R. Saitoh\inst{2}
        \and Yutaka Hirai\inst{3}
    }

    \institute{
        Department of Astronomy, Graduate School of Science, The University of Tokyo, 7-3-1 Hongo, Bunkyo-ku, Tokyo 113-0033, Japan
        \and Department of Planetology, Kobe University, Kobe, Hyogo 657-8501, Japan
        \and Department of Community Service and Science, Tohoku University of Community Service and Science, 3-5-1 Iimoriyama, Sakata, Yamagata 998-8580, Japan
    }

    \date{Received September 30, 20XX}
 
    \abstract
    {Binary populations are closely linked to the star formation process; however, their primordial properties can be changed by subsequent dynamical interactions within their natal clusters.}
    {The aim of this study is to clarify how different primordial binary populations affect the evolution of multiplicity and the global structure of forming star clusters.}
    {We investigate the dynamical evolution of primordial binaries during star cluster formation using self-consistent $N$-body/smoothed particle hydrodynamics simulations that follow the collapse of a molecular cloud to a star cluster. We systematically compare three star formation models: a close binary formation model (CB), a wide binary formation model (WB), and a single star formation model (SS).}
    {In CB and WB models, the multiplicity fraction decreases with time due to dynamical interactions. In particular, the fraction in the WB model drops to a level comparable to that in the SS model. The multiplicity fraction of high-mass stars is similarly high in all models, whereas only the CB model shows a relatively high fraction for low-mass stars. Due to the assumption of equal-mass binary formation, the CB and WB models exhibit an excess at $q=1$ in the mass-ratio distribution, while the SS model has no clear trend. Frequent few-body interactions generate distinct stellar populations inside and outside the cluster: the multiplicity fraction within the cluster is systematically higher, while mass functions in the outside have a shallower slope. Finally, stellar density profiles in the clusters are broadly similar among all models.
    }
    {The primordial binary population significantly affects the final binary properties, while having only a limited impact on their host cluster structures. Our results suggest that close binaries need to form at the star formation stage to reproduce the observed multiplicity fraction of low-mass stars and the excess of equal-mass binaries.
    }


    \maketitle
    \nolinenumbers

\section{Introduction}
A number of stars are found in binary or higher-order multiple systems with the multiplicity fraction increasing with primary mass.
Moreover, a wide range of theoretical and observational studies suggest that binary systems commonly emerge as a natural outcome of the star formation process \citep[see reviews by][]{Goodwin2007,Reipurth2014,Offner2023}.
In this context, binaries that form directly during the star formation phase are referred to as ``primordial'' binaries.
The statistical properties of primordial binaries, such as the multiplicity fraction and the orbital separation distribution, are expected to retain clear imprints of the physical conditions of star formation \citep[e.g.,][]{Duchene2013,Offner2023}.

Most stars are born in star clusters rather than isolated environments \citep{Lada2003}.
Primordial binaries can evolve dynamically via few-body interactions in clusters: the disruption and the dynamical formation of binaries can alter the multiplicity fraction and the orbital separation distribution, while partner exchange can change the mass ratios of binary systems \citep[e.g.,][]{Heggie1975,Hills1975b,Kroupa1995}.
As a result, the observed binary population reflects a complex interplay between binary formation and subsequent dynamical evolution \citep[e.g.,][]{Goodwin2007,Parker2014}.
Observational surveys support this picture.
For instance, \citet{Deacon2020} showed that the fraction of wide binaries (in the projected separation range $300$--$\SI{3000}{au}$) in open clusters is significantly lower than the field binary fraction \citep{Raghavan2010}, indicating that wide systems are preferentially disrupted in clustered environments due to dynamical interactions.
The Orion Nebula Cluster (ONC) also exhibits a low wide binary fraction relative to the field \citep{Reipurth2007,Jerabkova2019}.
In contrast, when focusing on close binaries, the binary fraction in the ONC is comparable to that in other star-forming regions \citep{Duchene2018}, suggesting that the binary formation process may be insensitive to the environment.
Moreover, no strong correlation is found between the close binary fraction and the age of clusters \citep{Kounkel2019}, implying that the primordial population of close binaries is largely preserved during cluster evolution.
Collectively, these observations suggest that the environmental dependence of binary populations differs between wide and close binary systems, reflecting their distinct responses to dynamical evolution in clustered environments.
However, observational constraints alone are insufficient to disentangle the imprint of binary formation from the effects of later dynamical interactions within forming star clusters.
This limitation motivates the use of numerical simulations that self-consistently follow star cluster formation from molecular clouds while computing the dynamical evolution of individual stars and binary systems.

Previous numerical studies based on pure $N$-body simulations, which adopt star clusters containing primordial binaries as initial conditions, have demonstrated that wide binaries are easily disrupted in high-density clusters \citep[e.g.,][]{Parker2009,Marks2012,Parker2014}.
The results highlight the need to explore earlier phases of star cluster evolution, in particular the cluster formation phase.
Several studies have carried out $N$-body simulations, including primordial binaries with initial conditions obtained by a hydrodynamical model or simulation \citep[e.g.,][]{Farias2017,Farias2019,Torniamenti2021}.
This approach allows for a more realistic description of the early dynamical evolution of star clusters.
However, these models do not self-consistently couple the dynamics of gas and stars during the cluster formation process.
Recently, \citet{Cournoyer-Cloutier2021,Cournoyer-Cloutier2024} performed simulations that couple magnetohydrodynamics with $N$-body stellar dynamics to follow the dynamical evolution of primordial binaries during star cluster formation.
By varying the mass and density of the parent molecular cloud, they showed that the overall binary fraction decreases during cluster formation, and that the properties of primordial binaries are more strongly affected in denser and more massive clouds.
In their framework, the properties of primordial binaries, such as the binary fraction and the orbital separation distribution, are prescribed on the basis of observations.
However, because the evolution of primordial binaries during star cluster formation is still not well constrained, it is essential to quantitatively compare the final properties of binaries resulting from different  primordial binary populations.

In this study, we investigate the dynamical evolution of primordial binaries during star cluster formation using smoothed particle hydrodynamics (SPH) coupled with $N$-body simulations. 
We focus on star cluster formation from high-density molecular clouds, where primordial binaries are expected to be strongly affected by dynamical interactions, and systematically compare different star formation prescriptions: close binary, wide binary, and single star formation models.
Section~\ref{sec:methods} describes the numerical methods and the star and binary formation models implemented.
In Section~\ref{sec:results}, we present the evolution of primordial binaries during cluster formation, while Section~\ref{sec:discussion} discusses the dynamics of high-mass stars.
Finally, Section~\ref{sec:conclusions} summarizes the main conclusions of this paper.

\section{Methods} \label{sec:methods}
\subsection{Numerical simulation}
We performed star cluster formation simulations starting from a molecular cloud as the initial condition.
The simulations were carried out using the \texttt{ASURA+BRIDGE} code \citep[SIRIUS project;][]{Hirai2021,Fujii2021b,Fujii2021a,Fujii2022b,Fujii2022a}, which is based on the $N$-body/SPH simulation code \texttt{ASURA} \citep{Saitoh2008,Saitoh2009}.
In this code, the BRIDGE scheme \citep{Fujii2007} couples the gas dynamics with a tree-based SPH method and the stellar dynamics with a high-order integration scheme, allowing us to follow the dynamical evolution of star clusters without gravitational softening.
The coupling between gas and stars is performed at a fixed Bridge timestep $\Delta t_{\mathrm{B}}$, at which star particles are kicked with the acceleration from gas potential.
In this study, we adopt a Bridge timestep of $\Delta t_{\mathrm{B}} = \SI{200}{yr}$.

For stellar dynamics in star clusters, \texttt{ASURA+BRIDGE} incorporates an $N$-body simulation code \texttt{PETAR} \citep{Wang2020b}, which employs the particle-particle particle-tree (P$^3$T) scheme \citep{Oshino2011,Iwasawa2015}.
The transition between the particle–particle (hard) and particle–tree (soft) parts is controlled by the changeover radii $r_{\mathrm{in}}$ and $r_{\mathrm{out}}$.
We adopt $r_{\mathrm{out}} = \SI{0.001}{pc}$ and $r_{\mathrm{in}} = r_{\mathrm{out}}/10$.
The time step for the soft part is set to $\Delta t_{\mathrm{soft}} = \Delta t_\mathrm{B}/1024$.
Thanks to its implementation of the slow-down algorithmic regularization (SDAR) scheme \citep{Wang2020a}, \texttt{PETAR} can accurately follow the orbital evolution of hard binaries and close encounters within  star clusters.
The radius $r_{\mathrm{bin}}$ used to identify multiple systems to which the SDAR scheme is applied is automatically determined in \texttt{PETAR}.
\texttt{PETAR} also incorporates stellar and binary evolution packages based on SSE/BSE \citep{Hurley2000,Hurley2002}; in this study, we adopted BSEEMP, updated by \citet{Tanikawa2020}.
For MPI and OpenMP parallelization, \texttt{PETAR} uses the framework for developing particle simulators \citep[FDPS;][]{Iwasawa2016}.

The feedback model from high-mass stars has been implemented in ASURA+BRIDGE code \citep{Fujii2021b}. For stars with masses above $8\,M_{\odot}$, the Str\"{o}mgren radius is calculated based on the local gas density and the ionizing photon counts per unit time obtained from OSTAR2002 model \citep{2003ApJS..146..417L}.
Gas particles located within the Str\"{o}mgren radius are heated to $10^{4}$\,K to mimic the formation of H{\sc ii} regions.
In addition, momentum injection due to radiation pressure and stellar winds is taken into account by the radial velocity kick to gas particles.
The kick velocity is given by
\begin{equation}
    \Delta v \propto \frac{Q \langle h\nu \rangle_i}{M_\mathrm{H_{II}} c} \Delta t,
\end{equation}
where $Q$ is the ionizing photon rate, $\langle h\nu \rangle_i = \SI{18.6}{eV}$ is the mean photon energy, $M_\mathrm{H_{II}}$ is the gas mass within the Str\"{o}mgren radius, $c$ is the speed of light, and $\Delta t$ is the timestep for integration.
The resulting kick velocities are typically below $\SI{100}{km.s^{-1}}$ \citep[see][for details]{Fujii2021b}.

\subsection{Star and binary formation}
\subsubsection{Star formation model} \label{sec:star_formation}
The star formation model in \texttt{ASURA+BRIDGE} is implemented as a stochastic (probabilistic) creation of single star particles when local gas conditions meet a set of prescribed criteria.
For each gas particle, the following conditions are evaluated:
(1) the local gas density exceeds a threshold density $n_\mathrm{th}$, (2) the gas temperature is below a threshold temperature $T_\mathrm{th}$, and (3) the local flow is converging ($\bm{\nabla \cdot v} < 0$).
If a gas particle satisfies all three criteria, it becomes eligible for star formation.
Then, we set the star formation probability
\begin{equation}
    p = \frac{m_\mathrm{gas}}{\langle m_* \rangle} \left[1 - \exp \left(-c_* \frac{\Delta t}{t_\mathrm{dyn}} \right)\right], 
\end{equation}
where $c_*, m_\mathrm{gas}, \langle m_* \rangle, \Delta t,$ and $t_\mathrm{dyn}$ are the dimensionless star formation efficiency, the mass of a gas particle, the averaged stellar mass in the given initial mass function (IMF), the time interval, and the local dynamical time, respectively.
If the probability exceeds the random number $R$ from 0 to 1 ($p > R$), a new star particle is created.
The mass of the new star is assigned randomly from the Kroupa mass function \citep{Kroupa2001} within the mass range $\SI{0.1}{M_\odot}$ to $\SI{150}{M_\odot}$.
The sampled stellar mass is accepted only if the total gas mass enclosed within a spherical search radius $r_\mathrm{search}$ centered on the candidate particle is greater than the sampled mass; otherwise the stellar mass is re-sampled from the IMF.
In this study, we adopt $n_\mathrm{th} = \SI{e5}{cm^{-3}}, T_\mathrm{th} = \SI{30}{K}, c_*=0.02$, and $r_\mathrm{search} = \SI{0.2}{pc}$.
Detailed descriptions are provided in \citet{Hirai2021}.

\subsubsection{Binary formation model} \label{s:binary_model}
Based on the above model, we developed a new module that forms binary stars instead of single stars.
The concept is that the positions and velocities of the two stars are determined from the two-body problem solution assuming that the binary follows an elliptical orbit.
The formation criteria for binaries are identical to those for single stars; that is, in this study, we assume that all newly formed systems are binaries.
The mass of the primary star $m_\mathrm{p}$ is stochastically sampled from the IMF.
Then, to determine the secondary mass $m_\mathrm{s}$ and the total binary mass $M$, a secondary-to-primary mass ratio $q$ is assigned.

Following the single star formation model, the position and velocity of the center of mass of the system $\bm{r}_\mathrm{com}, \bm{v}_\mathrm{com}$ are determined.
The orbital elements of the binary are then assigned stochastically.
Once a semi-major axis $a$ and eccentricity $e$ are given, the positions and velocities of the binary stars are determined as follows.
The mean anomaly $\theta_\mathrm{M}$ is uniformly distributed between $0$ and $2\pi$.
The eccentric anomaly $\theta_\mathrm{E}$ is obtained by solving Kepler's equation,
\begin{equation}
    \theta_\mathrm{M} = \theta_\mathrm{E} - e \sin \theta_\mathrm{E}.
\end{equation}
The separation between the two stars is given by
\begin{equation}
    r = a (1 - e \cos \theta_\mathrm{E}).
\end{equation}
Once $r$ and $\theta_\mathrm{E}$ are known, the position of the binary components in the orbital plane can be described.

The orbital plane orientation is specified by an arbitrary unit normal vector $\bm{n}$.
A unit vector pointing to the pericenter $\bm{e}_1$ is arbitrarily assigned any direction perpendicular to the normal vector $\bm{n}$, and the other unit vector $\bm{e}_2 = \bm{n}\times\bm{e}_1$ completes the orthogonal triad.
Using these unit vectors, the relative position vector $\bm{r}$ can be expressed as
\begin{equation}
    \bm{r} = r (\cos \theta_\mathrm{T} \bm{e}_1 + \sin \theta_\mathrm{T} \bm{e}_2),
\end{equation}
where $\theta_\mathrm{T}$ is the true anomaly related to $\theta_\mathrm{E}$ by
\begin{equation}
    \tan \left(\frac{\theta_\mathrm{T}}{2}\right) = \sqrt{\frac{1 + e}{1-e}} \tan \left(\frac{\theta_\mathrm{E}}{2}\right).
\end{equation}
Given the center-of-mass position $\bm{r}_\mathrm{com}$, the positions of the primary and secondary stars are then
\begin{align}
    \bm{r}_\mathrm{p} = \bm{r}_\mathrm{com} - \frac{m_\mathrm{s}}{M} \bm{r}, \\
    \bm{r}_\mathrm{s} = \bm{r}_\mathrm{com} + \frac{m_\mathrm{p}}{M} \bm{r}.
\end{align}
Similarly, the relative velocity $\bm{v}$ can be written as
\begin{equation}
    \bm{v} = \dot{\bm{r}} = \frac{GM}{j} \{-\sin \theta_\mathrm{T} \bm{e}_1 + (e + \cos \theta_\mathrm{T}) \bm{e}_2\},
\end{equation}
where $j = \sqrt{GMa(1-e^2)}$ is the specific angular momentum, and the velocities of the two stars are
\begin{align}
    \bm{v}_\mathrm{p} = \bm{v}_\mathrm{com} - \frac{m_\mathrm{s}}{M} \bm{v}, \\
    \bm{v}_\mathrm{s} = \bm{v}_\mathrm{com} + \frac{m_\mathrm{p}}{M} \bm{v}.
\end{align}
Through this procedure, each gas particle that satisfies the star formation criteria is converted into a binary system represented by two star particles.

In this study, the semi-major axis $a$ is drawn from a uniform distribution between $a_\mathrm{min}$ and $a_\mathrm{max}$, and the eccentricity $e$ is uniformly distributed between $0$ and $0.9$.
We explore two families of models that differ only in the adopted semi-major axis distribution: (i) a close binary formation model (CB) with $(a_\mathrm{min}, a_\mathrm{max}) = (1, 100)\,\si{au}$, and (ii) a wide binary formation model (WB) with $(a_\mathrm{min}, a_\mathrm{max}) = (100, 10^4)\,\si{au}$ (see Table~\ref{tab:model}).
The choice of the boundary between the CB and WB models is physically motivated by the hard-soft boundary of binaries in the cluster.
Hard binaries are defined by comparison with the average kinetic energy of stars in the stellar system.
If the binary is hard, it tends to evolve harder as a result of dynamical interactions with the other stars in the system \citep[e.g.,][]{Heggie1975,Hills1975a,Hills1975b}.
From the velocity dispersion of star clusters formed in our simulations without  binary formation, we adopted $\SI{100}{au}$ as the dividing scale between the CB and WB models.
For an easier comparison between the adopted models, we fix $q = 1$.

For comparison, we also performed the simulations with the single star formation model (SS) described in Section~\ref{sec:star_formation}.
In the binary formation models (CB and WB), each star formation event produces two star particles, whereas the SS model generates a single star particle.
As a result, for the same star formation efficiency $c_*$, the SS model forms stellar mass at roughly half the rate of the binary models.
To achieve a comparable amount of stellar mass, we adopted $c_* = 0.04$ for the SS model, twice the value used in the binary formation models.

\subsection{Initial condition} \label{s:ic}
The initial condition is an isolated, homogeneous spherical molecular cloud.
A turbulent velocity field was imposed on the cloud with a power spectrum proportional to $v^{-4}$, following \citet{Bate2003}.
We adopted a cloud mass of $M_\mathrm{cloud}=\num{5e3}\,M_\odot$, a cloud radius of $R_\mathrm{cloud}=\SI{2}{pc}$, an initial uniform temperature of $T_0=\SI{30}{K}$, an initial free-fall time of $t_{\mathrm{ff}} \simeq \SI{0.66}{Myr}$, and the initial virial ratio of $\alpha_\mathrm{vir}=|E_\mathrm{k}|/|E_\mathrm{p}|=1.0$, where $E_\mathrm{k}$ and $E_\mathrm{p}$ are kinetic and potential energy, respectively.

We adopted a gas particle mass of $m_\mathrm{gas}=0.1\,M_\odot$ in this study following \citet{Fujii2021b}.
In their study, two simulations using gas particle masses of $m_\mathrm{gas}=0.1$ and $0.01\,M_\odot$ were compared, and it was confirmed that the global properties of star cluster formation are not significantly affected by this choice of mass resolution.
The softening lengths of the gas and star particles are $\epsilon_\mathrm{g}=\SI{e4}{au}$ and $\epsilon_\mathrm{s}=\SI{0}{au}$, respectively.
The initial conditions are generated using the \texttt{molecular\_cloud} module in the Astrophysical Multi-purpose Software Environment \citep[AMUSE;][]{Portegies2009,Portegies2013,Pelupessy2013,Portegies2018,portegies_zwart_2023_8409512}.
These parameters are summarized in Table~\ref{tab:ic}.
We adopted the same random seed for the turbulent velocity fields for all the simulations.

To improve statistical robustness, particularly in high-mass stars, we performed three runs for each model (CB, WB, and SS) by varying the random seed used for sampling stellar masses in the star formation model.
This random seed alters the formation timing of massive stars, which can slightly change the timing of gas expulsion.
\begin{table}[ht!]
	\caption{Parameters of the binary formation models}
	\label{tab:model}
	\centering
	\begin{tabular}{ccc}
		\hline\hline
		  Model & $a_\mathrm{min}$\,(au) & $a_\mathrm{max}$\,(au) \\
		\hline
		CB & $1$    & $10^2$ \\
		WB & $10^2$ & $10^4$ \\
		\hline
	\end{tabular}
\end{table}

\begin{table}[ht!]
\caption{Parameters for the simulations}
\label{tab:ic}
\centering
\begin{tabular}{clc}
	\hline\hline
    Parameter & Description & Value \\
    \hline
	  $M_\mathrm{cloud}$       & Cloud mass                & $5\times10^3\,M_\odot$ \\
    $R_\mathrm{cloud}$       & Cloud radius              & $\SI{2}{pc}$ \\
    $T_0$                    & Initial temperature       & $\SI{30}{K}$ \\
    $\rho_0$                 & Initial density           & $\SI{e-20}{g.cm^{-3}}$ \\
    $t_{\mathrm{ff},0}$      & Initial free-fall time    & $\SI{0.66}{Myr}$ \\
    $\alpha_\mathrm{vir}$    & Initial virial ratio      & $1.0$ \\
    $m_\mathrm{gas}$         & Gas particle mass         & $0.1\,M_\odot$ \\
    $\epsilon_\mathrm{gas}$  & Softening length for gas  & $\SI{e4}{au}$ \\
    $\epsilon_\mathrm{star}$ & Softening length for star & $\SI{0}{au}$ \\
	\hline
\end{tabular}
\end{table}

\section{Results} \label{sec:results}
\subsection{Star cluster formation} \label{sec:cluster_formation}
\begin{figure*}[htbp!]
    \centering
    \includegraphics[width=0.85\linewidth]{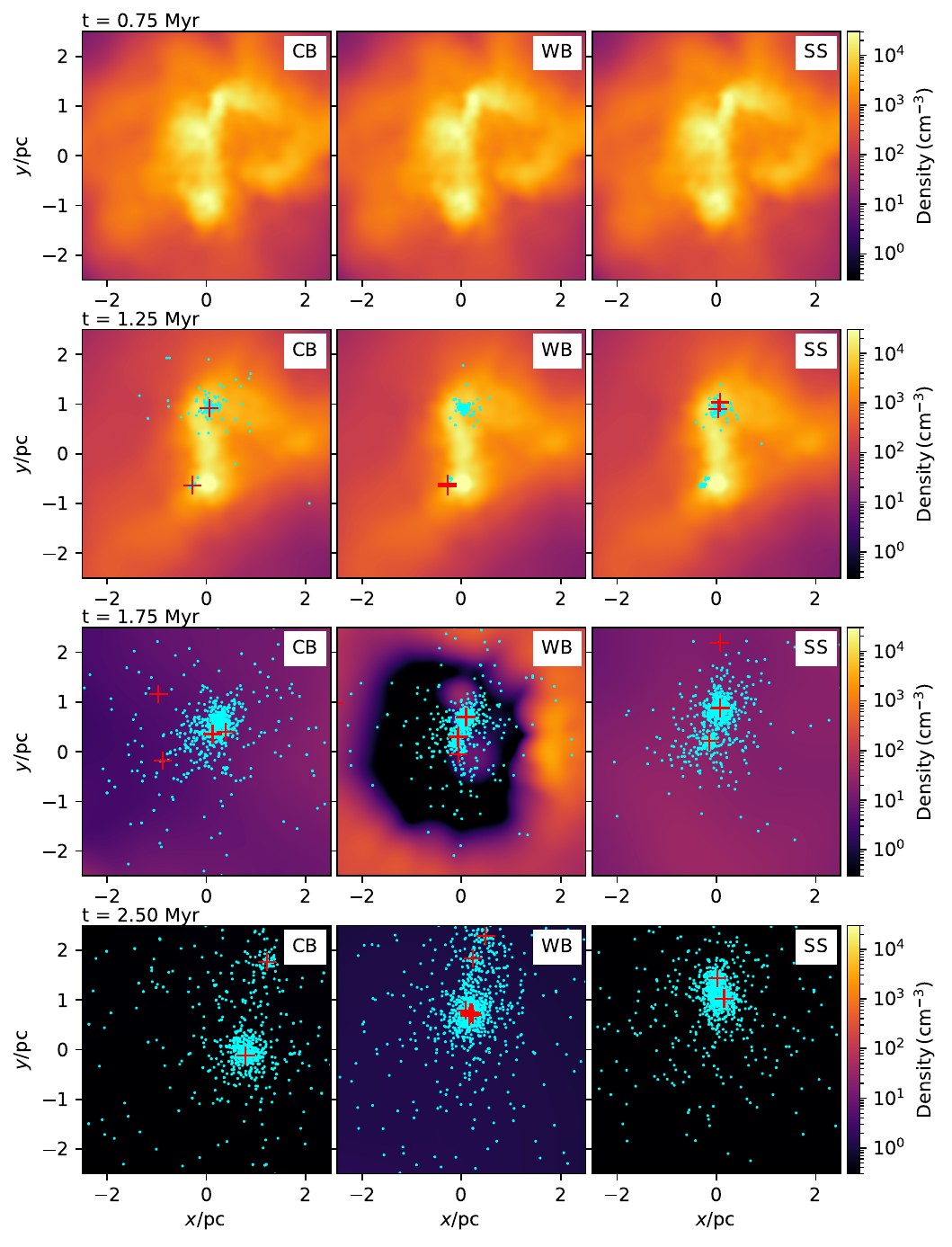}
    \caption{
    Snapshots of the star cluster formation simulations at four representative epochs.
    From left to right, the panels show the CB, WB, and SS models.
    In each panel, the gas density slice in the $z=0$ plane is shown in color, while cyan dots indicate the positions of stars.
    Red crosses mark the positions of high-mass stars.
    }
    \label{f:2d_map}
\end{figure*}
Fig.~\ref{f:2d_map} presents snapshots of star cluster formation simulations with the three models: the close binary model (CB), the wide binary model (WB), and the single star model (SS).
The colors show the gas density in the $z=0$ plane, visualized by using the Pynbody package \citep{pynbody}.
Fig.~\ref{f:2d_map} shows the representative results of one of the three runs.
The initial condition is a spherical cloud with a total gas mass of $M = 5000\,M_\odot$ and a radius $R=\SI{2}{pc}$, corresponding to the free-fall time of $t_\mathrm{ff} \simeq \SI{0.66}{Myr}$ (see Section\,\ref{s:ic} and Table\,\ref{tab:ic}).
Star formation begins at about one free-fall time.
The results presented in this paper cover the evolution up to $t=\SI{2.5}{Myr} \sim 4 t_\mathrm{ff}$.

Since all models share identical initial conditions, the overall distribution of the particles is nearly indistinguishable among the three models during the early phase of star formation.
However, after $t\sim \SI{1.75}{Myr}$, the spatial distributions of stars and gas begin to diverge gradually.
This divergence arises from the variations in the timing of high-mass star formation between the models. In the case of this random seed, high-mass stars happened to form earlier in the CB and SS models, and the resulting feedback leads to more rapid gas expulsion compared to the WB model.

\begin{figure}
	\includegraphics[width=\columnwidth]{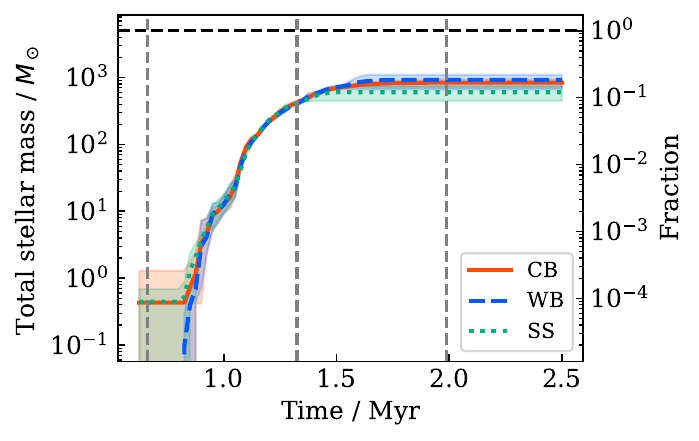}
    \caption{
    Time evolution of the total stellar mass.
    The red-solid, blue-dashed, and green-dotted lines correspond to the CB, WB, and SS models, respectively.
    Thick lines indicate the fraction averaged over simulations with different random seeds, while thin lines represent the corresponding minimum and maximum values among those seeds.
    The gray vertical lines indicate $1$, $2$, and $3\times t_\mathrm{ff}$.
    }
    \label{f:mass_evo}
\end{figure}

\begin{figure}
	\includegraphics[width=0.9\columnwidth]{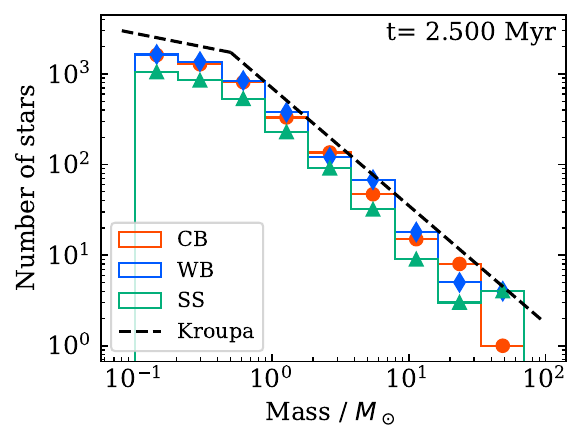}
    \caption{
    Stellar mass functions at $t=\SI{2.5}{Myr}$.
    The CB, WB, and SS models are shown by red circles, blue diamonds, and green triangles, respectively.
    The histograms are constructed by combining the simulations with three different random seeds.
    The black dashed line indicates the Kroupa IMF adopted in the star formation model.
    }
    \label{f:mf}
\end{figure}

The time evolution of the total stellar mass in each model is shown in Fig.~\ref{f:mass_evo}.
The overall growth of stellar mass is very similar across all the models. At a later time, the average total stellar mass of the SS model is slightly lower than the others. This is also caused by the timing of first high-mass star formation, which suppresses subsequent star formation.
Fig.~\ref{f:mf} shows the stellar mass functions at $t=\SI{2.5}{Myr}$ for all three models.
The mass functions of the CB and WB models are nearly indistinguishable, and all models broadly reproduce the \citet{Kroupa2001} IMF prescribed in the star formation model (see Section~\ref{sec:star_formation}).

\subsection{Binary formation and evolution} \label{s:binary}
Binary and higher-order multiple systems were identified in each snapshot.
First, for every stellar particle, we searched for its nearest neighbor.
Pairs that are mutual nearest neighbors (i.e., particle $j$ is the nearest neighbor of $i$ and $i$ is the nearest neighbor of $j$) were collected as binary candidates.
For each candidate pair, we computed the two-body total energy,
\begin{equation}
    E = \frac{1}{2} v_\mathrm{rel}^2 - G\frac{M}{r},
\end{equation}
where $M = m_i + m_j$ is the total mass of the pair, $v_\mathrm{rel} = |\bm{v}_i - \bm{v}_j|$ the relative velocity, and $r=|\bm{r}_i - \bm{r}_j|$ the separation of the pair.
Pairs with $E<0$ were classified as gravitationally bound binaries.
To identify hierarchical multiples, each confirmed binary was replaced by a composite (virtual) particle characterized by the binary's total mass, center-of-mass position, and center-of-mass velocity;
the same nearest-neighbor + binding test was then applied to the enlarged particle set.
Following this recursive procedure, we detected higher-order multiple systems.

\begin{figure}
    \centering
    \includegraphics[width=0.9\columnwidth]{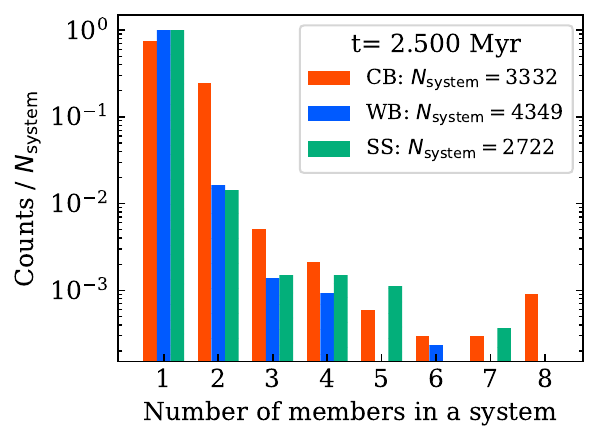}
    \caption{
        Histograms of the numbers of single (1), binary (2), triple (3), and higher-order systems at the end of the simulations.
        The red, blue, and green bars correspond to the CB, WB, and SS models, respectively.
        The histograms are constructed by combining the simulations with different random seeds.
    }
    \label{f:hist_Nmember}
\end{figure}

Fig.~\ref{f:hist_Nmember} presents histograms of the number of single, binary, triple, and higher-order systems at the end of the simulations.
The legend of Fig.~\ref{f:hist_Nmember} shows the total number of stellar systems. As discussed in Section~\ref{sec:cluster_formation}, early formation of high-mass stars suppressed subsequent star formation, resulting in a relatively low total stellar mass and number of stars in the SS model (see also Figs.~\ref{f:mass_evo} and \ref{f:mf}). This leads to a smaller total number of systems in the SS model compared with the CB and WB models.
Even in the SS model, some binaries are produced dynamically during cluster formation; the final multiplicity fraction is of the order of $1\%$.
Despite the initial $100\%$ binary fraction in the CB and WB models, a lot of primordial binaries are disrupted during cluster formation, so that single stars constitute the dominant population at the final epoch.
Disruption is particularly frequent in the WB model: wide binaries are preferentially destroyed, resulting in a final multiplicity comparable to that of the SS model ($\sim 1\%$).
Triples and higher-order systems are present but rare in all models.

\begin{figure}
    \centering
    \includegraphics[width=0.9\columnwidth]{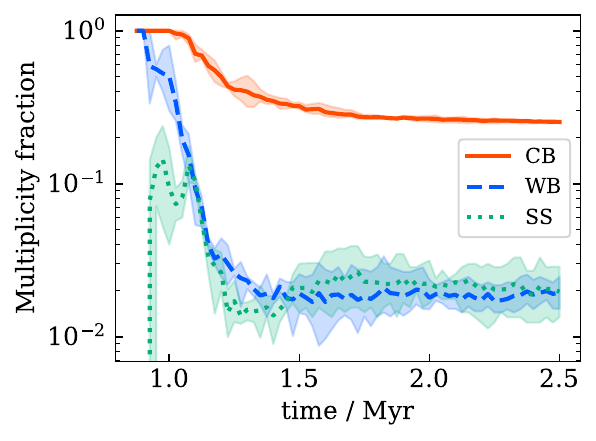}
    \caption{
        Time evolution of the multiplicity fraction.
        The CB, WB, and SS models are shown by the red-solid, blue-dashed, and green-dotted lines, respectively.
        Thick lines indicate the fraction averaged over simulations with different random seeds, while thin lines represent the corresponding minimum and maximum values among those seeds.
    }
    \label{f:MF_evo}
\end{figure}

Fig.~\ref{f:MF_evo} displays the time evolution of the multiplicity fraction.
The multiplicity fraction is defined as the proportion of multiple systems relative to the total number of systems, i.e.,
\begin{equation}
    MF = \frac{B+T+\cdots}{S+B+T+\cdots},
\end{equation}
where $S, B,$ and $T$ are the number of single stars, binaries, and triples, respectively.
As expected, the multiplicity fractions in the CB and WB models are initially 100\% by construction, but these decline sharply as star formation proceeds and dynamical interactions become frequent.
The WB model shows a rapid decline.
Most primordial wide binaries are destroyed, and the multiplicity fraction falls toward values similar to the SS model.
By contrast, the CB model retains a certain fraction of binaries, with the multiplicity fraction leveling off at tens of percent.

\begin{figure}
	\includegraphics[width=\linewidth]{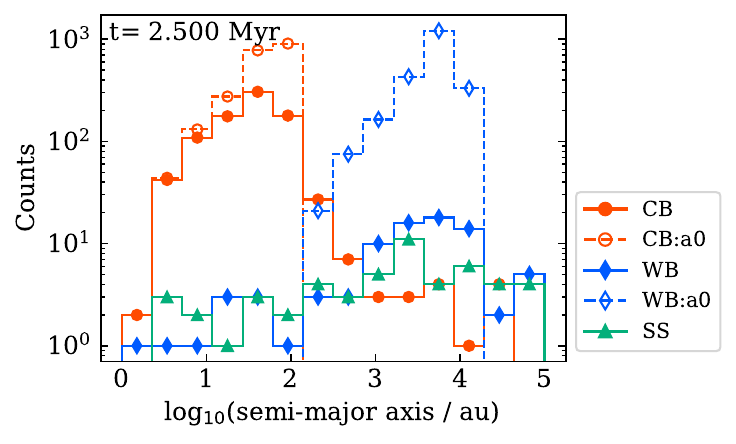}
    \caption{
    Semi-major axis distributions of binaries at the final epoch.
    The CB, WB, and SS models are shown in red circles, blue diamonds, and green triangles, respectively.
    The dashed lines with unfilled symbols in the same colors indicate the semi-major axis distributions of all ``primordial'' binaries formed up to the end of the simulation in each model.
    The histograms are constructed by combining the simulations with different random seeds.
    }
    \label{f:separation}
\end{figure}

Fig.~\ref{f:separation} shows the semi-major axis distributions at the final epoch (solid lines), together with those of all primordial binaries formed up to the end of the simulation in the binary-formation models CB and WB (dashed lines).
The figure reveals a clear distinction between the `soft' and `hard' binaries: systems with separations of $\gtrsim \SI{100}{au}$ are typically soft and are readily disrupted by subsequent encounters (particularly prominent in the WB model), whereas binaries with separations below $\sim \SI{100}{au}$ are hard and have small cross sections, resulting in a high probability of survival.
This trend is consistent with recent simulations of star cluster formation that include primordial binaries \citep{Cournoyer-Cloutier2024}.
In the SS model, only a small number of binaries are formed through few-body encounters.

\begin{figure*}
	\includegraphics[width=\linewidth]{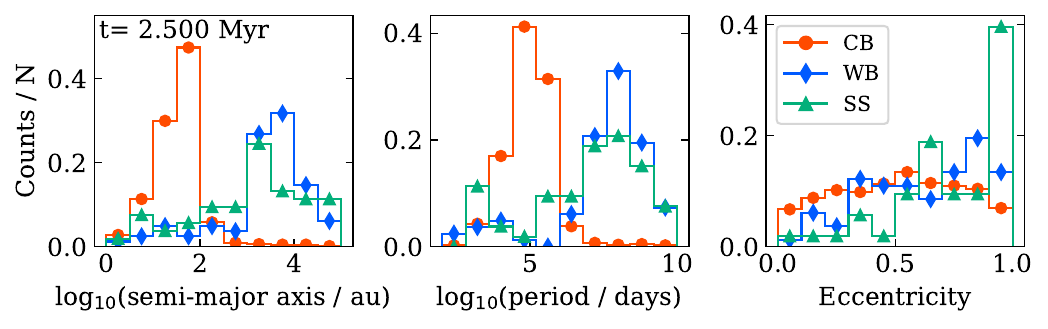}
    \caption{
    Distributions of binary parameters at the end of the simulations.
    From left to right, the panels show the distributions of semi-major axis, orbital period, and eccentricity.
    The red circles, blue diamonds, and green triangles correspond to the CB, WB, and SS models, respectively.
    The histograms are constructed by combining the simulations with different random seeds.
    }
    \label{f:ape_distribution}
\end{figure*}

Fig.~\ref{f:ape_distribution} shows the distributions of binary parameters—semi-major axis, orbital period, and eccentricity—at the final epoch of the simulations.
In the SS model, dynamically formed binaries exhibit semi-major axis distributions similar to those of the WB model, predominantly at wide separation ($\gg \SI{100}{au}$).
\citet{Torniamenti2021} reported a similar result: soft binaries, whose binding energy is comparable to or below the cluster's energy scale, are continuously formed and disrupted through gravitational interactions.
The eccentricity distribution of the CB model remains nearly flat, showing little difference from the primordial distribution prescribed in the binary formation model.
By contrast, the distributions in the WB and SS models show an upward trend toward higher eccentricities.
This trend is qualitatively consistent with a thermal distribution $p(e) = 2e$ \citep[e.g.,][]{Jeans1919}, suggesting that the binaries in the WB and SS models have frequently experienced dynamical interactions during cluster formation.

\begin{figure}
    \centering
    \includegraphics[width=0.9\columnwidth]{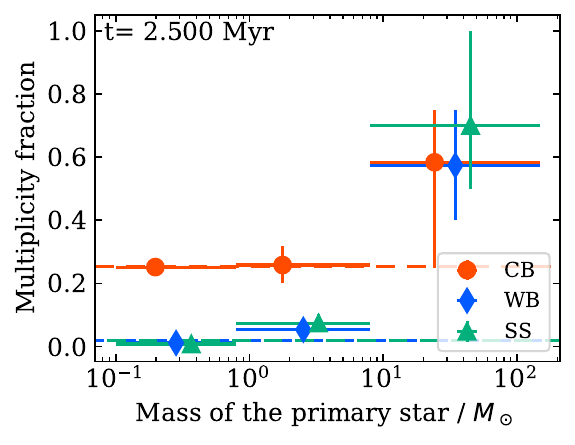}
    \caption{
        Multiplicity fractions as a function of the mass of the primary star at the end of the simulation.
        Red circles, blue diamonds, and green triangles show the mean multiplicity fraction over simulations with different random seeds for the CB, WB, and SS models, respectively.
        The vertical error bars indicate the range between the minimum and maximum values among the three simulations, while the horizontal error bars represent the mass ranges of each bin.
        The symbols are slightly offset from the bin centers for clarity.
        Horizontal dashed lines indicate the overall multiplicity fraction obtained by combining the results from three simulations for each model.
    }
    \label{f:mf_vs_mp}
\end{figure}

Fig.~\ref{f:mf_vs_mp} shows the multiplicity fraction as a function of the primary mass at the final epoch;
the multiplicity calculated over all systems is indicated by dashed lines.
As already shown in Fig.~\ref{f:MF_evo}, the overall multiplicity is the highest in the CB model, while the WB and SS models exhibit only a few percent.
However, the overall multiplicity is strongly biased by the large number of low-mass stars.
The horizontal axis in Fig.~\ref{f:mf_vs_mp} is divided into three primary-mass bins: low-mass ($m_\mathrm{p} < 0.8\,M_\odot$), intermediate-mass ($0.8 \leq m_\mathrm{p}/M_\odot < 8$), and high-mass ($m_\mathrm{p} \geq 8\,M_\odot$).
The multiplicity fraction of high-mass primaries shows large scatter due to the small number; however, compared to the lower-mass primaries, it looks higher across the three models.
Then, we find a positive multiplicity trend with primary mass.
Notably, although the CB and WB models assume a $100$\% binary fraction at formation, nearly $40$\% of high-mass binaries are disrupted on average.
One possible reason for this low multiplicity fraction is the inclusion of stars that have been dynamically ejected from the cluster.

\begin{figure*}
    \sidecaption
    \centering
    \includegraphics[width=12cm]{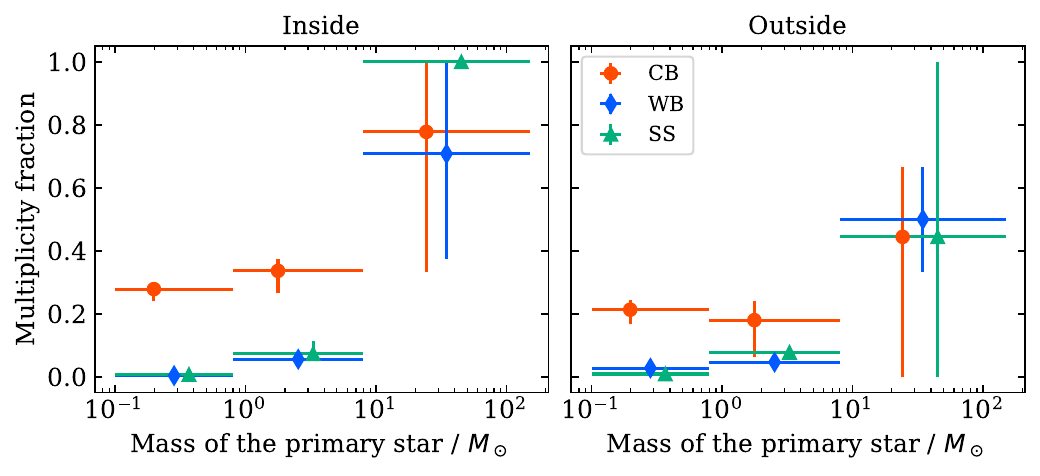}
    \caption{
        Multiplicity fractions as a function of primary mass for stars located inside (left) and outside (right) the cluster at the end of the simulation.
        These are plotted with the same representation as in Fig.~\ref{f:mf_vs_mp}.
    }
    \label{f:multiplicity_inout}
\end{figure*}

Fig.~\ref{f:multiplicity_inout} shows the multiplicity fraction as a function of the primary mass for stars located inside and outside the cluster at the end of the simulation.
For simplicity, we define the inside of the cluster as the region within $\SI{2}{pc}$ of the peak stellar density.
The multiplicity fraction of high-mass stars shows a spatial variation which is higher in the inner region than in the outer region of the clusters.
In particular, in the SS model, all high-mass stars inside the cluster have a companion in all three simulations with different random seeds, i.e., the multiplicity fraction is unity.
This spatial trend is consistent with observations for the Orion Nebula Cluster \citep[ONC;][]{Kohler2006}, which reported that the binary fraction of stars with masses $>2\,M_\odot$ is lower in the periphery than in the cluster center.
The ONC is a young cluster hosting several high-mass stars, still embedded in the natal molecular cloud, making it a suitable target for comparison with our simulations.
The detailed disruption process of high-mass binaries is discussed in Section~\ref{sec:high-mass}.

The multiplicity fraction of high-mass stars is similarly high in all models, suggesting that the imprint of the (high-mass) star formation process is largely erased by subsequent dynamical evolution.
In contrast, the multiplicity fraction of low-mass stars shows significant differences among the models and is particularly low in the SS model.
It is consistent with previous studies \citep[e.g.,][]{Cournoyer-Cloutier2021}, which showed that forming stars purely as single systems is insufficient to reproduce the observed multiplicity fractions.
Furthermore, the similarly low multiplicity fraction in the WB model suggests that forming only wide primordial binaries is also insufficient, implying that a significant population of close primordial binaries is likely required.
Such systems are dynamically hard and thus capable of surviving dynamical processing within a forming cluster.

\begin{figure}
    \centering
    \includegraphics[width=0.9\linewidth]{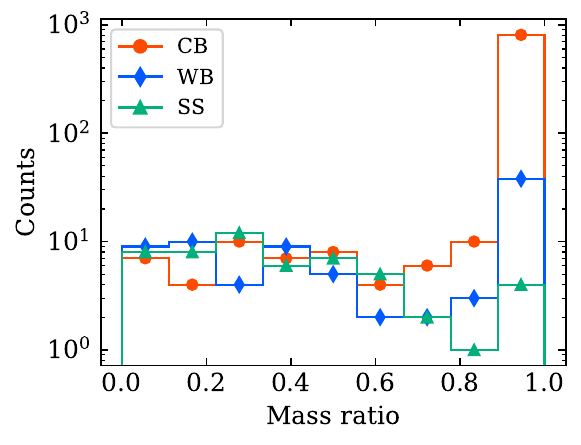}
    \caption{
    Mass ratio distributions of binaries at the end of the simulations.
    Red circles, blue diamonds, and green triangles correspond to the CB, WB, and SS models, respectively.
    The histograms are constructed by combining the simulations with different random seeds.
    }
    \label{f:q_dist}
\end{figure}

Fig.~\ref{f:q_dist} shows the distribution of binary mass ratios, defined as $q = m_{\mathrm{s}}/m_{\mathrm{p}}$.
The mass-ratio distribution in the SS model is nearly flat, without any clear trend.
In contrast, the CB and WB models exhibit approximately flat distributions over a wide range of $q$ but show a clear excess around $q \simeq 1$.
This feature reflects the binary formation prescription described in Section~\ref{s:binary_model}, in which all binaries are assumed to form as equal-mass ``twin'' binaries.
In the CB model, most binaries remain concentrated near $q=1$, indicating that a substantial fraction of the primordial binaries survive the subsequent cluster formation phase.
On the other hand, in the WB model, about half of the 82 binaries are primordial, whereas the other ones are considered to have experienced partner exchange or formed dynamically from single stars.

Observationally, an excess of twins has been reported across a wide range of stellar populations, including solar-type stars \citep{Tokovinin2000,Raghavan2010}, OB stars \citep{Moe2017}, and young stars \citep{Kounkel2019}, especially in close binaries.
As described above, it is difficult for dynamical interactions to produce an excess of twins.
These results therefore suggest that equal-mass binaries are likely born via the star formation process.

\section{Discussion} \label{sec:discussion}
\subsection{Dynamical evolution of high-mass binaries} \label{sec:high-mass}
As shown in Fig.~\ref{f:multiplicity_inout}, the multiplicity fraction as a function of primary mass shows a spatial dependence, especially for high-mass stars.
The fraction is lower in the outer region compared to the inner region of the cluster.
One possible interpretation is that (high-mass) binaries are disrupted in the cluster center, and their components are subsequently ejected into the outer regions.

\begin{figure}
    \centering
    \includegraphics[width=\linewidth]{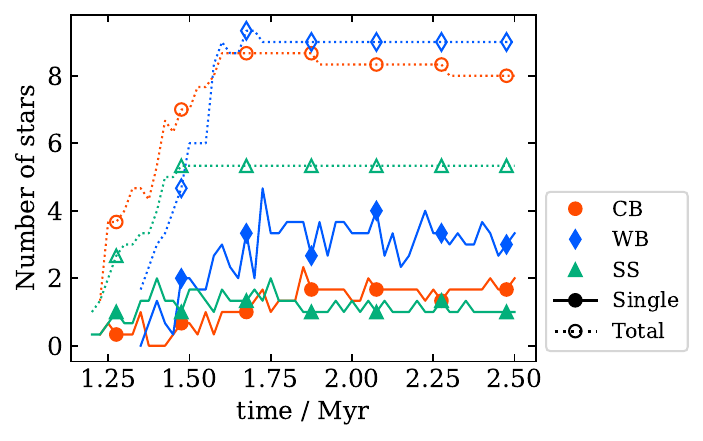}
    \caption{
    Time evolution of the number of high-mass stars.
    Solid lines show the number of singles, while dotted lines show the total number of high-mass stars.
    Red circles, blue diamonds, and green triangles correspond to the CB, WB, and SS models, respectively.
    These are averaged over simulations with different random seeds.
    }
    \label{f:N_Time_massive}
\end{figure}

We first examine the time evolution of the number of high-mass stars ($>8\,M_{\odot}$) in Fig.~\ref{f:N_Time_massive}.
Solid and dotted lines show the number of single and total high-mass stars, respectively.
The difference between them corresponds to the number of high-mass stars in multiple systems.
The total number increases as star formation proceeds until $\sim \SI{1.7}{Myr}$, and the number of singles is maintained to be a few per single run for all models.
This means that, in the SS model, newly formed single high-mass stars quickly acquire companions through few-body interactions, and in the CB and WB models, newly formed binaries are not disrupted efficiently.
The WB model shows a slightly larger number of single high-mass stars than the CB model.
Note that the total number of high-mass stars does not increase monotonically.
This is not caused by mass loss through stellar evolution but is instead due to mergers between high-mass stars.
We traced such merger events using simulation snapshots and detected 5, 4, and 1 events in the CB, WB, and SS models, respectively.
Since the merging components are mostly binary members, these mergers can also increase the number of single high-mass stars. The smaller number of mergers in the SS model is likely due to the difference in the star formation model and/or the smaller number of high-mass stars formed.

Binary disruption can be enhanced in dense cluster environments and during subcluster mergers, which can promote frequent few-body interactions \citep[e.g.,][]{Fujii2012,Rantala2024,Cournoyer-Cloutier2024_ClusterAssembly}.
Mergers of subclusters with gas can also enhance star and binary formation \citep[e.g.,][]{Fujii2022a,Karam2025}.
In this work, a few subclusters are formed within $r<\SI{2}{pc}$, but each subcluster has only a few high-mass stars, and the number of them is too small to robustly characterize such enhancement.

\begin{figure}
    \centering
    \includegraphics[width=0.9\linewidth]{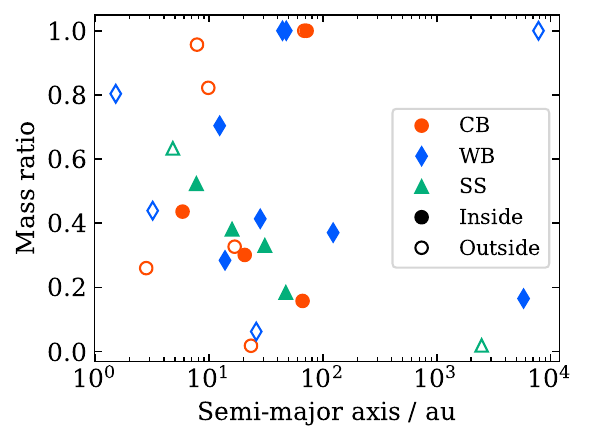}
    \caption{
    Distribution of the semi-major axis and mass ratio of high-mass binaries at the end of the simulations.
    Red circles, blue diamonds, and green triangles correspond to the CB, WB, and SS models, respectively.
    Filled and open symbols represent binaries in the inner and outer regions of the cluster, respectively.
    These are constructed by combining the simulations with different random seeds.
    }
    \label{f:high_mass_a_q}
\end{figure}

Since high-mass binaries are not efficiently disrupted, we next investigate whether the binaries preserve their original pairs or have undergone partner exchange.
Fig.~\ref{f:high_mass_a_q} shows the distributions of the semi-major axis and the mass ratio of high-mass binaries at the end of the simulations. For comparison, we also showed the results of the SS model.
The mass ratio of high-mass binaries is broadly distributed, indicating that a large fraction of high-mass twin binaries formed in the CB and WB models have experienced partner exchange through few-body encounters.
Although the number of samples is not enough, no clear differences are seen in either the semi-major axes or mass ratios between high-mass binaries in the inner and outer regions of the cluster, which are represented by filled and open symbols, respectively.

The broader mass ratio distribution of high-mass binaries contrasts with that seen for binaries over the full mass range (Fig.~\ref{f:q_dist}), in which at least half of them retain mass ratios close to unity.
This is likely because high-mass stars preferentially form in dense gas-rich regions and subsequently sink toward the cluster center through mass segregation, where few-body interactions occur more frequently.

Next, we focus on the kinematics of high-mass stars located in the outer regions ($>2$\,pc) of the cluster.
Stars dynamically ejected from clusters are known as runaway and walkaway stars ($\gtrsim \SI{10}{km.s^{-1}}$), both of which have been observed in the ONC \citep[e.g.,][]{McBride2019,Schoettler2020}.
We counted high-mass systems located outside the cluster with radial velocities exceeding $\SI{10}{km.s^{-1}}$: the numbers are 3, 4, and 1 for the CB, WB, and SS models, respectively.
This suggests that some high-mass stars are dynamically ejected from the cluster center via few-body encounters.
Although the sample size is too small for a statistically robust conclusion, binary formation models appear to produce a larger fraction of runaway stars.

\begin{figure*}
    \sidecaption
    \centering
    \includegraphics[width=12cm]{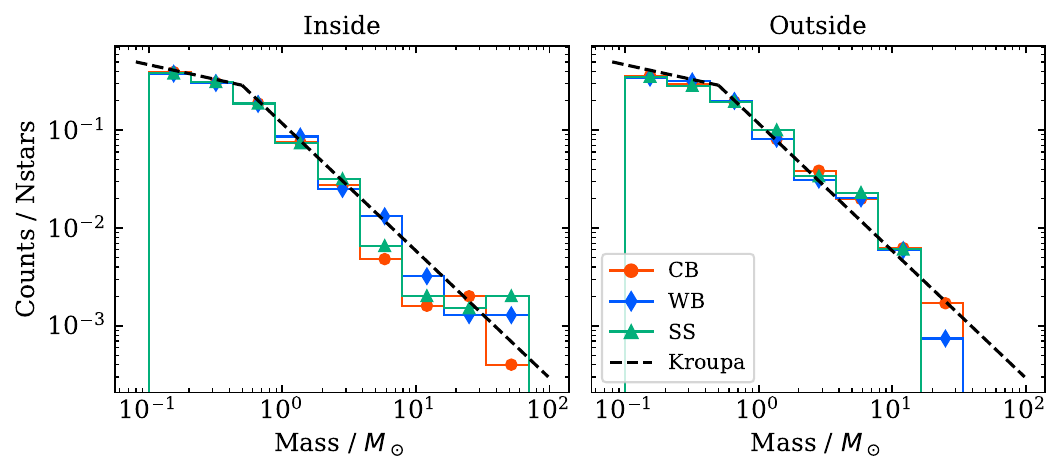}
    \caption{
    Stellar mass functions for stars located inside (left) and outside (right) the cluster at the end of the simulation.
    These are constructed by combining the simulations with different random seeds.
    Red circles, blue diamonds, and green triangles correspond to the CB, WB, and SS models, respectively.
    }
    \label{f:mf_inout}
\end{figure*}

Theoretical studies have shown that high-mass stars can be preferentially ejected from cluster centers through few-body interactions during dense cluster formation \citep[e.g.,][]{Fujii2011,Fujii2012,Oh2016}.
Thus, we examine the stellar mass functions for stars located inside and outside the cluster in Fig.~\ref{f:mf_inout}.
High-mass stars sink toward the cluster center due to mass segregation. In the central region, dynamical interactions with the most massive binaries eject the other moderately massive stars into the outer regions.
As a result, the most massive stars remain located in the cluster center, while the moderately massive stars become depleted there.
Conversely, they are enhanced in the outer region, leading to a shallower mass function.
To statistically test whether the mass function differs between the inside and outside of the cluster, we performed a two-sample Kolmogorov–Smirnov test for each model.
The resulting p-values are 0.013, 0.057, and 0.024 for the CB, WB, and SS models, respectively.
The null hypothesis that the inside and outside mass functions are drawn from the same distribution is rejected at the $5$\% significance level for the CB and SS models, while the WB model shows marginal evidence for a difference.
These results support the interpretation that mass segregation leads to a systematic difference between the mass functions inside and outside the cluster.
Mass segregation has been observed in young clusters such as the ONC \citep[e.g.,][]{Hillenbrand1998,Allison2009_ONC}.
Numerical studies have shown that high-mass stars can become mass-segregated within a few Myr through the early dynamical evolution of clusters (e.g., subcluster mergers: \citealt{McMillan2007,Moeckel2009}; subvirial and fractal clusters: \citealt{Allison2009}).
Our results are consistent with these studies, showing that mass segregation develops during the early phase of cluster formation.

\subsection{Impact of binaries on cluster density}
In this subsection, we focus on the impact of different binary populations on the global structure of the forming cluster.
Binaries are known to play an important role in the dynamical evolution of star clusters, as they act as an energy source through gravitational interactions with surrounding stars \citep[e.g.,][]{Heggie1975}.
Binaries with intermediate hardness can efficiently supply energy and therefore suppress core collapse \citep[e.g.,][]{McMillan1990,2009PASJ...61..721T}.
Here, we compare the properties of clusters formed in different star formation models.

\begin{figure*}
    \sidecaption
	\includegraphics[width=12cm]{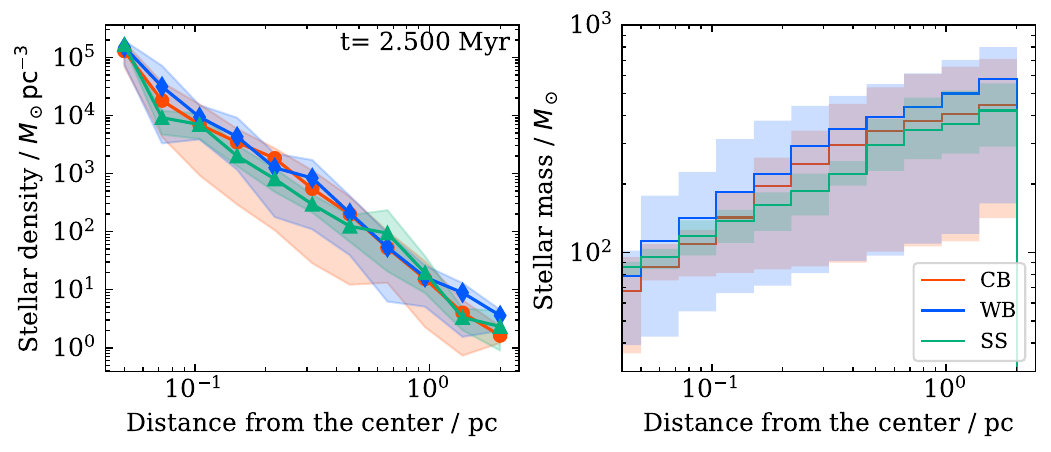}
    \caption{
    Stellar number density profiles (left) and cumulative stellar mass distributions (right) at the end of the simulation.
    The horizontal axis shows the distance from the cluster center, defined as the position of the peak stellar density.
    Red circles, blue diamonds, and green triangles correspond to the CB, WB, and SS models, respectively.
    Thick lines indicate the averaged profiles over simulations with different random seeds, while thin lines represent the corresponding minimum and maximum values among those seeds.
    }
    \label{f:density_profile}
\end{figure*}

The left panel of Fig.~\ref{f:density_profile} shows the stellar number density profiles measured in spherical shells centered on the peak of the stellar density, while the right panel presents the cumulative stellar mass distributions at the end of the simulation.
Radial profiles were derived using logarithmically spaced bins between $0.05$ and $\SI{2}{pc}$, and the innermost bin ($0$--$\SI{0.05}{pc}$) contains $>10$ stars even in the smallest case.
These radial profiles of the three models are broadly similar, indicating that the global cluster structure is only weakly affected by the adopted primordial binary population.
During the cluster formation phase, the gravitational potential of the gas is dominant, and binary heating through few-body interactions is expected to play a minor role in the global cluster dynamics.
Therefore, the overall cluster dynamics remains largely insensitive to the existence of primordial binaries.
This picture is consistent with \citet{Cournoyer-Cloutier2023}, who performed simulations of star cluster formation and found that the dynamical evolution of young embedded clusters with masses ($\lesssim 10^3\,M_\odot$) is driven by the gravitational potential of the star-forming region rather than by few-body encounters.

Whether the present similarity in cluster structure among the CB, WB, and SS models would persist during subsequent dynamical evolution under the gas-free condition is beyond the scope of this study and is left for future work.
\citet{Torniamenti2021} performed $N$-body simulations of young cluster evolution for $\SI{10}{Myr}$ using realistic initial conditions derived from hydrodynamical simulations, which provide a good analogy for the near-future, gas-free state of the clusters formed in this study.
They found that binaries enhance the expansion of the clusters; in this study, since the CB model retains a much larger population of close, hard binaries than the WB and SS models, a similar mechanism could make the clusters in the CB model larger than those of the other models.
On the other hand, \citet{Amiri2026} followed the dynamical evolution of star clusters with initial substructure for $\SI{100}{Myr}$ using $N$-body simulations and found that the primordial hard binary fraction (0\% vs. 10\%) does not significantly affect the resulting degree of mass segregation by the end of the simulations.
Although the clusters in their simulations are a few times more massive than those in our models,
this suggests that, at least in terms of the degree of mass segregation, our models may not diverge significantly even in the subsequent dynamical evolution stage.

\subsection{Caveats}
In this study, we adopted a simple prescription for the primordial binary population to focus on the effect of their semi-major axis on the cluster formation under controlled conditions.
While this approach allows for a clearer interpretation of the results, care must be taken in making direct comparisons with observations.

First, we assumed that all primordial binaries have equal-mass companions ($q=1$).
However, observational studies suggest that binaries span a broad range of mass ratios \citep[e.g.,][]{Moe2017}, thereby leading to systematically lower binding energies than in our simulations.
Binaries with lower binding energies are more fragile, and therefore, the overall multiplicity fraction would likely decrease.
Nevertheless, as shown in Fig.~\ref{f:high_mass_a_q}, the mass ratios of high-mass binaries have various values through subsequent dynamical interactions, despite the assumption of equal-mass binary formation.

Second, we assumed that the distribution of the semi-major axis is independent of the primary mass.
However, observations have shown that binaries with more massive primaries tend to have smaller median separations \citep[see e.g., Figs. 2 and 3 in][]{Offner2023}.
If such a mass-dependent distribution were included, higher-mass binaries would have higher binding energies and smaller cross sections, which could suppress their disruption and affect the mass dependence of the multiplicity fraction.

A quantitative comparison with observations using more realistic primordial binary distributions will be explored in future work \citep[see also][]{Cournoyer-Cloutier2021,Cournoyer-Cloutier2024}.

\section{Conclusions} \label{sec:conclusions}
In this study, we have investigated the dynamical evolution of primordial binaries during star cluster formation by means of self-consistent $N$-body/SPH simulations that follow the collapse of a molecular cloud into a star cluster.
We systematically compared three star formation models: a close binary formation model (CB), a wide binary formation model (WB), and a single star formation model (SS), in order to assess how the properties of primordial binaries affect the resulting multiplicity and cluster structure.

In binary formation models, the multiplicity fraction decreased with time in both the CB and WB models.
The multiplicity fraction of the WB model dropped to $\sim 1$\%, which was comparable to that of the SS model. 
On the other hand, the CB model maintained a relatively high multiplicity fraction of $20$ -- $30$\%.
This is because the wide binaries in our model are soft for the formed star clusters, and therefore, they are easily broken up via dynamical encounters.
The decline of the multiplicity fraction saturated after $\sim \SI{1.5}{Myr}$, corresponding to two to three free-fall times of the parent cloud.

The dependence of the multiplicity fraction on the mass of the primary star shows that all models maintain a high multiplicity fraction for high-mass stars.
In contrast, consistent with previous studies \citep[e.g.,][]{Cournoyer-Cloutier2021}, the SS model produces a multiplicity fraction for low-mass stars that is significantly lower than observed values.
Despite the assumption of $100$\% binary formation, a similar trend is also found in the WB model, suggesting that close binaries are already formed during the star formation stage.

Our results show that reproducing an excess of equal-mass binaries requires the presence of a non-negligible population of primordial twins.
In the SS model lacking such primordial systems, the mass-ratio distribution became nearly flat at the end of the simulations.
This suggests that the observed twin excess, particularly among close binaries, is already established at the star formation phase rather than being produced by subsequent dynamical evolution.

Higher-mass stars preferentially sink toward the cluster center through mass segregation, where frequent dynamical interactions can eject some of them into the outer regions of the cluster.
As a result, distinct stellar populations emerge inside and outside the cluster.
The multiplicity fraction is systematically higher in the inner region than in the outer region, while the mass function in the outer region is relatively shallower than that in the inner region.

The impact of the primordial binary population on the structures of the forming star cluster is limited.
The stellar density profiles in the cluster are broadly similar among three models, which is likely because the gas potential dominates at the cluster formation phase.

Our conclusions are based on simulations that adopt a dense molecular cloud as an initial condition, corresponding to the regime of young massive cluster formation.
In lower-density environments, where close encounters are less frequent, quantitative results may differ.
However, this would primarily shift the boundary between dynamically hard and soft binaries, while the qualitative trend that closer binaries are likely to retain the imprint of the star formation process is expected to remain unchanged.
Exploring a broader range of cloud densities, masses, and primordial binary populations, as well as including additional physics such as the magnetic field, will be important topics for future work.

\begin{acknowledgements}
We thank the anonymous referee for insightful comments.
This research was supported by KAKENHI Grant Numbers 23K22530, 21K03614, 21K03633, 22K03688, 24K07095, 25K01046, 25H00664, 25K17434, 26K00744, and 26K07153.
\end{acknowledgements}

\bibliographystyle{aa}
\bibliography{references.bib}

@ARTICLE{McMillan1990,
       author = {{McMillan}, Steve and {Hut}, Piet and {Makino}, Junichiro},
        title = "{Star Cluster Evolution with Primordial Binaries. I. A Comparative Study}",
      journal = {\apj},
         year = 1990,
        month = oct,
       volume = {362},
        pages = {522},
          doi = {10.1086/169289},
       adsurl = {https://ui.adsabs.harvard.edu/abs/1990ApJ...362..522M}
}

@ARTICLE{2003ApJS..146..417L,
       author = {{Lanz}, Thierry and {Hubeny}, Ivan},
        title = "{A Grid of Non-LTE Line-blanketed Model Atmospheres of O-Type Stars}",
      journal = {\apjs},
         year = 2003,
        month = jun,
       volume = {146},
       number = {2},
        pages = {417-441},
          doi = {10.1086/374373},
archivePrefix = {arXiv},
       eprint = {astro-ph/0210157},
 primaryClass = {astro-ph},
       adsurl = {https://ui.adsabs.harvard.edu/abs/2003ApJS..146..417L}
}

@ARTICLE{2009PASJ...61..721T,
       author = {{Tanikawa}, Ataru and {Fukushige}, Toshiyuki},
        title = "{Effects of Hardness of Primordial Binaries on the Evolution of Star Clusters}",
      journal = {\pasj},
         year = 2009,
        month = aug,
       volume = {61},
        pages = {721},
          doi = {10.1093/pasj/61.4.721},
archivePrefix = {arXiv},
       eprint = {1005.2237},
 primaryClass = {astro-ph.GA},
       adsurl = {https://ui.adsabs.harvard.edu/abs/2009PASJ...61..721T}
}

@ARTICLE{Jeans1919,
       author = {{Jeans}, J.~H.},
        title = "{The origin of binary systems}",
      journal = {\mnras},
         year = 1919,
        month = apr,
       volume = {79},
        pages = {408},
          doi = {10.1093/mnras/79.6.408},
       adsurl = {https://ui.adsabs.harvard.edu/abs/1919MNRAS..79..408J}
}

@ARTICLE{Heggie1975,
       author = {{Heggie}, D.~C.},
        title = "{Binary evolution in stellar dynamics.}",
      journal = {\mnras},
         year = 1975,
        month = dec,
       volume = {173},
        pages = {729-787},
          doi = {10.1093/mnras/173.3.729},
       adsurl = {https://ui.adsabs.harvard.edu/abs/1975MNRAS.173..729H}
}

@ARTICLE{Hills1975b,
       author = {{Hills}, J.~G.},
        title = "{Encounters between binary and single stars and their effect on the dynamical evolution of stellar systems.}",
      journal = {\aj},
         year = 1975,
        month = oct,
       volume = {80},
        pages = {809-825},
          doi = {10.1086/111815},
       adsurl = {https://ui.adsabs.harvard.edu/abs/1975AJ.....80..809H}
}

@ARTICLE{Hills1975a,
       author = {{Hills}, J.~G.},
        title = "{Effect of binary stars on the dynamical evolution of stellar clusters. II. Analytic evolutionary models.}",
      journal = {\aj},
         year = 1975,
        month = dec,
       volume = {80},
        pages = {1075-1080},
          doi = {10.1086/111842},
       adsurl = {https://ui.adsabs.harvard.edu/abs/1975AJ.....80.1075H}
}

@ARTICLE{Kroupa1995,
       author = {{Kroupa}, Pavel},
        title = "{Inverse dynamical population synthesis and star formation}",
      journal = {\mnras},
         year = 1995,
        month = dec,
       volume = {277},
        pages = {1491},
          doi = {10.1093/mnras/277.4.1491},
archivePrefix = {arXiv},
       eprint = {astro-ph/9508117},
 primaryClass = {astro-ph},
       adsurl = {https://ui.adsabs.harvard.edu/abs/1995MNRAS.277.1491K}
}

@ARTICLE{Hillenbrand1998,
       author = {{Hillenbrand}, Lynne A. and {Hartmann}, Lee W.},
        title = "{A Preliminary Study of the Orion Nebula Cluster Structure and Dynamics}",
      journal = {\apj},
         year = 1998,
        month = jan,
       volume = {492},
       number = {2},
        pages = {540-553},
          doi = {10.1086/305076},
       adsurl = {https://ui.adsabs.harvard.edu/abs/1998ApJ...492..540H}
}

@ARTICLE{Tokovinin2000,
       author = {{Tokovinin}, A.~A.},
        title = "{On the origin of binaries with twin components}",
      journal = {\aap},
         year = 2000,
        month = aug,
       volume = {360},
        pages = {997-1002},
       adsurl = {https://ui.adsabs.harvard.edu/abs/2000A&A...360..997T}
}

@ARTICLE{Kroupa2001,
       author = {{Kroupa}, Pavel},
        title = "{On the variation of the initial mass function}",
      journal = {\mnras},
         year = 2001,
        month = apr,
       volume = {322},
       number = {2},
        pages = {231-246},
          doi = {10.1046/j.1365-8711.2001.04022.x},
archivePrefix = {arXiv},
       eprint = {astro-ph/0009005},
 primaryClass = {astro-ph},
       adsurl = {https://ui.adsabs.harvard.edu/abs/2001MNRAS.322..231K}
}

@ARTICLE{Bate2003,
       author = {{Bonnell}, Ian A. and {Bate}, Matthew R. and {Vine}, Stephen G.},
        title = "{The hierarchical formation of a stellar cluster}",
      journal = {\mnras},
         year = 2003,
        month = aug,
       volume = {343},
       number = {2},
        pages = {413-418},
          doi = {10.1046/j.1365-8711.2003.06687.x},
archivePrefix = {arXiv},
       eprint = {astro-ph/0305082},
 primaryClass = {astro-ph},
       adsurl = {https://ui.adsabs.harvard.edu/abs/2003MNRAS.343..413B}
}

@ARTICLE{Lada2003,
       author = {{Lada}, Charles J. and {Lada}, Elizabeth A.},
        title = "{Embedded Clusters in Molecular Clouds}",
      journal = {\araa},
         year = 2003,
        month = jan,
       volume = {41},
        pages = {57-115},
          doi = {10.1146/annurev.astro.41.011802.094844},
archivePrefix = {arXiv},
       eprint = {astro-ph/0301540},
 primaryClass = {astro-ph},
       adsurl = {https://ui.adsabs.harvard.edu/abs/2003ARA&A..41...57L}
}

@ARTICLE{Kohler2006,
       author = {{K{\"o}hler}, R. and {Petr-Gotzens}, M.~G. and {McCaughrean}, M.~J. and {Bouvier}, J. and {Duch{\^e}ne}, G. and {Quirrenbach}, A. and {Zinnecker}, H.},
        title = "{Binary stars in the Orion Nebula Cluster}",
      journal = {\aap},
         year = 2006,
        month = nov,
       volume = {458},
       number = {2},
        pages = {461-476},
          doi = {10.1051/0004-6361:20054561},
archivePrefix = {arXiv},
       eprint = {astro-ph/0607670},
 primaryClass = {astro-ph},
       adsurl = {https://ui.adsabs.harvard.edu/abs/2006A&A...458..461K}
}

@ARTICLE{Fujii2007,
       author = {{Fujii}, Michiko and {Iwasawa}, Masaki and {Funato}, Yoko and {Makino}, Junichiro},
        title = "{BRIDGE: A Direct-Tree Hybrid N-Body Algorithm for Fully Self-Consistent Simulations of Star Clusters and Their Parent Galaxies}",
      journal = {\pasj},
         year = 2007,
        month = dec,
       volume = {59},
        pages = {1095},
          doi = {10.1093/pasj/59.6.1095},
archivePrefix = {arXiv},
       eprint = {0706.2059},
 primaryClass = {astro-ph},
       adsurl = {https://ui.adsabs.harvard.edu/abs/2007PASJ...59.1095F}
}

@INPROCEEDINGS{Goodwin2007,
       author = {{Goodwin}, S.~P. and {Kroupa}, P. and {Goodman}, A. and {Burkert}, A.},
        title = "{The Fragmentation of Cores and the Initial Binary Population}",
    booktitle = {Protostars and Planets V},
         year = 2007,
       editor = {{Reipurth}, Bo and {Jewitt}, David and {Keil}, Klaus},
        month = jan,
        pages = {133},
          doi = {10.48550/arXiv.astro-ph/0603233},
archivePrefix = {arXiv},
       eprint = {astro-ph/0603233},
 primaryClass = {astro-ph},
       adsurl = {https://ui.adsabs.harvard.edu/abs/2007prpl.conf..133G}
}

@ARTICLE{Reipurth2007,
       author = {{Reipurth}, Bo and {Guimar{\~a}es}, Marcelo M. and {Connelley}, Michael S. and {Bally}, John},
        title = "{Visual Binaries in the Orion Nebula Cluster}",
      journal = {\aj},
         year = 2007,
        month = dec,
       volume = {134},
       number = {6},
        pages = {2272-2285},
          doi = {10.1086/523596},
archivePrefix = {arXiv},
       eprint = {0709.3824},
 primaryClass = {astro-ph},
       adsurl = {https://ui.adsabs.harvard.edu/abs/2007AJ....134.2272R}
}

@ARTICLE{McMillan2007,
       author = {{McMillan}, Stephen L.~W. and {Vesperini}, Enrico and {Portegies Zwart}, Simon F.},
        title = "{A Dynamical Origin for Early Mass Segregation in Young Star Clusters}",
      journal = {\apjl},
         year = 2007,
        month = jan,
       volume = {655},
       number = {1},
        pages = {L45-L49},
          doi = {10.1086/511763},
archivePrefix = {arXiv},
       eprint = {astro-ph/0609515},
 primaryClass = {astro-ph},
       adsurl = {https://ui.adsabs.harvard.edu/abs/2007ApJ...655L..45M}
}

@ARTICLE{Saitoh2008,
       author = {{Saitoh}, Takayuki R. and {Daisaka}, Hiroshi and {Kokubo}, Eiichiro and {Makino}, Junichiro and {Okamoto}, Takashi and {Tomisaka}, Kohji and {Wada}, Keiichi and {Yoshida}, Naoki},
        title = "{Toward First-Principle Simulations of Galaxy Formation: I. How Should We Choose Star-Formation Criteria in High-Resolution Simulations of Disk Galaxies?}",
      journal = {\pasj},
         year = 2008,
        month = aug,
       volume = {60},
       number = {4},
        pages = {667-681},
          doi = {10.1093/pasj/60.4.667},
archivePrefix = {arXiv},
       eprint = {0802.0961},
 primaryClass = {astro-ph},
       adsurl = {https://ui.adsabs.harvard.edu/abs/2008PASJ...60..667S}
}

@ARTICLE{Saitoh2009,
       author = {{Saitoh}, Takayuki R. and {Daisaka}, Hiroshi and {Kokubo}, Eiichiro and {Makino}, Junichiro and {Okamoto}, Takashi and {Tomisaka}, Kohji and {Wada}, Keiichi and {Yoshida}, Naoki},
        title = "{Toward First-Principle Simulations of Galaxy Formation: II. Shock-Induced Starburst at a Collision Interface during the First Encounter of Interacting Galaxies}",
      journal = {\pasj},
         year = 2009,
        month = jun,
       volume = {61},
        pages = {481},
          doi = {10.1093/pasj/61.3.481},
archivePrefix = {arXiv},
       eprint = {0805.0167},
 primaryClass = {astro-ph},
       adsurl = {https://ui.adsabs.harvard.edu/abs/2009PASJ...61..481S}
}

@ARTICLE{Parker2009,
       author = {{Parker}, Richard J. and {Goodwin}, Simon P. and {Kroupa}, Pavel and {Kouwenhoven}, M.~B.~N.},
        title = "{Do binaries in clusters form in the same way as in the field?}",
      journal = {\mnras},
         year = 2009,
        month = aug,
       volume = {397},
       number = {3},
        pages = {1577-1586},
          doi = {10.1111/j.1365-2966.2009.15032.x},
archivePrefix = {arXiv},
       eprint = {0905.2140},
 primaryClass = {astro-ph.GA},
       adsurl = {https://ui.adsabs.harvard.edu/abs/2009MNRAS.397.1577P}
}

@ARTICLE{Allison2009,
       author = {{Allison}, Richard J. and {Goodwin}, Simon P. and {Parker}, Richard J. and {de Grijs}, Richard and {Portegies Zwart}, Simon F. and {Kouwenhoven}, M.~B.~N.},
        title = "{Dynamical Mass Segregation on a Very Short Timescale}",
      journal = {\apjl},
         year = 2009,
        month = aug,
       volume = {700},
       number = {2},
        pages = {L99-L103},
          doi = {10.1088/0004-637X/700/2/L99},
archivePrefix = {arXiv},
       eprint = {0906.4806},
 primaryClass = {astro-ph.GA},
       adsurl = {https://ui.adsabs.harvard.edu/abs/2009ApJ...700L..99A}
}

@ARTICLE{Allison2009_ONC,
       author = {{Allison}, Richard J. and {Goodwin}, Simon P. and {Parker}, Richard J. and {Portegies Zwart}, Simon F. and {de Grijs}, Richard and {Kouwenhoven}, M.~B.~N.},
        title = "{Using the minimum spanning tree to trace mass segregation}",
      journal = {\mnras},
         year = 2009,
        month = may,
       volume = {395},
       number = {3},
        pages = {1449-1454},
          doi = {10.1111/j.1365-2966.2009.14508.x},
archivePrefix = {arXiv},
       eprint = {0901.2047},
 primaryClass = {astro-ph.GA},
       adsurl = {https://ui.adsabs.harvard.edu/abs/2009MNRAS.395.1449A}
}

@ARTICLE{Moeckel2009,
       author = {{Moeckel}, Nickolas and {Bonnell}, Ian A.},
        title = "{Does subcluster merging accelerate mass segregation in local clusters?}",
      journal = {\mnras},
         year = 2009,
        month = dec,
       volume = {400},
       number = {2},
        pages = {657-664},
          doi = {10.1111/j.1365-2966.2009.15499.x},
archivePrefix = {arXiv},
       eprint = {0908.0253},
 primaryClass = {astro-ph.SR},
       adsurl = {https://ui.adsabs.harvard.edu/abs/2009MNRAS.400..657M}
}

@ARTICLE{Raghavan2010,
       author = {{Raghavan}, Deepak and {McAlister}, Harold A. and {Henry}, Todd J. and {Latham}, David W. and {Marcy}, Geoffrey W. and {Mason}, Brian D. and {Gies}, Douglas R. and {White}, Russel J. and {ten Brummelaar}, Theo A.},
        title = "{A Survey of Stellar Families: Multiplicity of Solar-type Stars}",
      journal = {\apjs},
         year = 2010,
        month = sep,
       volume = {190},
       number = {1},
        pages = {1-42},
          doi = {10.1088/0067-0049/190/1/1},
archivePrefix = {arXiv},
       eprint = {1007.0414},
 primaryClass = {astro-ph.SR},
       adsurl = {https://ui.adsabs.harvard.edu/abs/2010ApJS..190....1R}
}

@ARTICLE{Oshino2011,
       author = {{Oshino}, Shoichi and {Funato}, Yoko and {Makino}, Junichiro},
        title = "{Particle-Particle Particle-Tree: A Direct-Tree Hybrid Scheme for Collisional N-Body Simulations}",
      journal = {\pasj},
         year = 2011,
        month = aug,
       volume = {63},
        pages = {881},
          doi = {10.1093/pasj/63.4.881},
archivePrefix = {arXiv},
       eprint = {1101.5504},
 primaryClass = {astro-ph.EP},
       adsurl = {https://ui.adsabs.harvard.edu/abs/2011PASJ...63..881O}
}

@ARTICLE{Fujii2011,
       author = {{Fujii}, Michiko S. and {Portegies Zwart}, Simon},
        title = "{The Origin of OB Runaway Stars}",
      journal = {Science},
         year = 2011,
        month = dec,
       volume = {334},
       number = {6061},
        pages = {1380},
          doi = {10.1126/science.1211927},
archivePrefix = {arXiv},
       eprint = {1111.3644},
 primaryClass = {astro-ph.GA},
       adsurl = {https://ui.adsabs.harvard.edu/abs/2011Sci...334.1380F}
}

@ARTICLE{Fujii2012,
       author = {{Fujii}, M.~S. and {Saitoh}, T.~R. and {Portegies Zwart}, S.~F.},
        title = "{The Formation of Young Dense Star Clusters through Mergers}",
      journal = {\apj},
         year = 2012,
        month = jul,
       volume = {753},
       number = {1},
          eid = {85},
        pages = {85},
          doi = {10.1088/0004-637X/753/1/85},
archivePrefix = {arXiv},
       eprint = {1205.1434},
 primaryClass = {astro-ph.GA},
       adsurl = {https://ui.adsabs.harvard.edu/abs/2012ApJ...753...85F}
}

@ARTICLE{Marks2012,
       author = {{Marks}, M. and {Kroupa}, P.},
        title = "{Inverse dynamical population synthesis. Constraining the initial conditions of young stellar clusters by studying their binary populations}",
      journal = {\aap},
         year = 2012,
        month = jul,
       volume = {543},
          eid = {A8},
        pages = {A8},
          doi = {10.1051/0004-6361/201118231},
archivePrefix = {arXiv},
       eprint = {1205.1508},
 primaryClass = {astro-ph.GA},
       adsurl = {https://ui.adsabs.harvard.edu/abs/2012A&A...543A...8M}
}

@ARTICLE{Duchene2013,
       author = {{Duch{\^e}ne}, Gaspard and {Kraus}, Adam},
        title = "{Stellar Multiplicity}",
      journal = {\araa},
         year = 2013,
        month = aug,
       volume = {51},
       number = {1},
        pages = {269-310},
          doi = {10.1146/annurev-astro-081710-102602},
archivePrefix = {arXiv},
       eprint = {1303.3028},
 primaryClass = {astro-ph.SR},
       adsurl = {https://ui.adsabs.harvard.edu/abs/2013ARA&A..51..269D}
}

@INPROCEEDINGS{Reipurth2014,
       author = {{Reipurth}, B. and {Clarke}, C.~J. and {Boss}, A.~P. and {Goodwin}, S.~P. and {Rodr{\'\i}guez}, L.~F. and {Stassun}, K.~G. and {Tokovinin}, A. and {Zinnecker}, H.},
        title = "{Multiplicity in Early Stellar Evolution}",
    booktitle = {Protostars and Planets VI},
         year = 2014,
       editor = {{Beuther}, Henrik and {Klessen}, Ralf S. and {Dullemond}, Cornelis P. and {Henning}, Thomas},
        month = jan,
        pages = {267-290},
          doi = {10.2458/azu_uapress_9780816531240-ch012},
archivePrefix = {arXiv},
       eprint = {1403.1907},
 primaryClass = {astro-ph.SR},
       adsurl = {https://ui.adsabs.harvard.edu/abs/2014prpl.conf..267R}
}

@ARTICLE{Parker2014,
       author = {{Parker}, Richard J. and {Meyer}, Michael R.},
        title = "{Binaries in the field: fossils of the star formation process?}",
      journal = {\mnras},
         year = 2014,
        month = aug,
       volume = {442},
       number = {4},
        pages = {3722-3736},
          doi = {10.1093/mnras/stu1101},
archivePrefix = {arXiv},
       eprint = {1406.0844},
 primaryClass = {astro-ph.SR},
       adsurl = {https://ui.adsabs.harvard.edu/abs/2014MNRAS.442.3722P}
}

@ARTICLE{Iwasawa2015,
       author = {{Iwasawa}, Masaki and {Portegies Zwart}, Simon and {Makino}, Junichiro},
        title = "{GPU-enabled particle-particle particle-tree scheme for simulating dense stellar cluster system}",
      journal = {Computational Astrophysics and Cosmology},
         year = 2015,
        month = jul,
       volume = {2},
          eid = {6},
        pages = {6},
          doi = {10.1186/s40668-015-0010-1},
archivePrefix = {arXiv},
       eprint = {1506.04553},
 primaryClass = {astro-ph.IM},
       adsurl = {https://ui.adsabs.harvard.edu/abs/2015ComAC...2....6I}
}

@ARTICLE{Iwasawa2016,
       author = {{Iwasawa}, Masaki and {Tanikawa}, Ataru and {Hosono}, Natsuki and {Nitadori}, Keigo and {Muranushi}, Takayuki and {Makino}, Junichiro},
        title = "{Implementation and performance of FDPS: a framework for developing parallel particle simulation codes}",
      journal = {\pasj},
         year = 2016,
        month = aug,
       volume = {68},
       number = {4},
          eid = {54},
        pages = {54},
          doi = {10.1093/pasj/psw053},
archivePrefix = {arXiv},
       eprint = {1601.03138},
 primaryClass = {astro-ph.IM},
       adsurl = {https://ui.adsabs.harvard.edu/abs/2016PASJ...68...54I}
}

@ARTICLE{Oh2016,
       author = {{Oh}, Seungkyung and {Kroupa}, Pavel},
        title = "{Dynamical ejections of massive stars from young star clusters under diverse initial conditions}",
      journal = {\aap},
         year = 2016,
        month = may,
       volume = {590},
          eid = {A107},
        pages = {A107},
          doi = {10.1051/0004-6361/201628233},
archivePrefix = {arXiv},
       eprint = {1604.00006},
 primaryClass = {astro-ph.GA},
       adsurl = {https://ui.adsabs.harvard.edu/abs/2016A&A...590A.107O}
}

@ARTICLE{Moe2017,
       author = {{Moe}, Maxwell and {Di Stefano}, Rosanne},
        title = "{Mind Your Ps and Qs: The Interrelation between Period (P) and Mass-ratio (Q) Distributions of Binary Stars}",
      journal = {\apjs},
         year = 2017,
        month = jun,
       volume = {230},
       number = {2},
          eid = {15},
        pages = {15},
          doi = {10.3847/1538-4365/aa6fb6},
archivePrefix = {arXiv},
       eprint = {1606.05347},
 primaryClass = {astro-ph.SR},
       adsurl = {https://ui.adsabs.harvard.edu/abs/2017ApJS..230...15M}
}

@ARTICLE{Farias2017,
       author = {{Farias}, Juan P. and {Tan}, Jonathan C. and {Chatterjee}, Sourav},
        title = "{Star Cluster Formation from Turbulent Clumps. I. The Fast Formation Limit}",
      journal = {\apj},
         year = 2017,
        month = apr,
       volume = {838},
       number = {2},
          eid = {116},
        pages = {116},
          doi = {10.3847/1538-4357/aa63f6},
archivePrefix = {arXiv},
       eprint = {1701.00701},
 primaryClass = {astro-ph.GA},
       adsurl = {https://ui.adsabs.harvard.edu/abs/2017ApJ...838..116F}
}

@ARTICLE{Duchene2018,
       author = {{Duch{\^e}ne}, G. and {Lacour}, S. and {Moraux}, E. and {Goodwin}, S. and {Bouvier}, J.},
        title = "{Is stellar multiplicity universal? Tight stellar binaries in the Orion nebula Cluster}",
      journal = {\mnras},
         year = 2018,
        month = aug,
       volume = {478},
       number = {2},
        pages = {1825-1836},
          doi = {10.1093/mnras/sty1180},
archivePrefix = {arXiv},
       eprint = {1805.00965},
 primaryClass = {astro-ph.SR},
       adsurl = {https://ui.adsabs.harvard.edu/abs/2018MNRAS.478.1825D}
}

@ARTICLE{Kounkel2019,
       author = {{Kounkel}, Marina and {Covey}, Kevin and {Moe}, Maxwell and {Kratter}, Kaitlin M. and {Su{\'a}rez}, Genaro and {Stassun}, Keivan G. and {Rom{\'a}n-Z{\'u}{\~n}iga}, Carlos and {Hernandez}, Jesus and {Kim}, Jinyoung Serena and {Pe{\~n}a Ram{\'\i}rez}, Karla and et al.},
        title = "{Close Companions around Young Stars}",
      journal = {\aj},
         year = 2019,
        month = may,
       volume = {157},
       number = {5},
          eid = {196},
        pages = {196},
          doi = {10.3847/1538-3881/ab13b1},
archivePrefix = {arXiv},
       eprint = {1903.10523},
 primaryClass = {astro-ph.SR},
       adsurl = {https://ui.adsabs.harvard.edu/abs/2019AJ....157..196K}
}

@ARTICLE{Jerabkova2019,
       author = {{Jerabkova}, Tereza and {Beccari}, Giacomo and {Boffin}, Henri M.~J. and {Petr-Gotzens}, Monika G. and {Manara}, Carlo F. and {Prada Moroni}, Pier Giorgio and {Tognelli}, Emanuele and {Degl'Innocenti}, Scilla},
        title = "{When the tale comes true: multiple populations and wide binaries in the Orion Nebula Cluster}",
      journal = {\aap},
         year = 2019,
        month = jul,
       volume = {627},
          eid = {A57},
        pages = {A57},
          doi = {10.1051/0004-6361/201935016},
archivePrefix = {arXiv},
       eprint = {1905.06974},
 primaryClass = {astro-ph.SR},
       adsurl = {https://ui.adsabs.harvard.edu/abs/2019A&A...627A..57J}
}

@ARTICLE{McBride2019,
       author = {{McBride}, Aidan and {Kounkel}, Marina},
        title = "{Runaway Young Stars near the Orion Nebula}",
      journal = {\apj},
         year = 2019,
        month = oct,
       volume = {884},
       number = {1},
          eid = {6},
        pages = {6},
          doi = {10.3847/1538-4357/ab3df9},
archivePrefix = {arXiv},
       eprint = {1908.07550},
 primaryClass = {astro-ph.SR},
       adsurl = {https://ui.adsabs.harvard.edu/abs/2019ApJ...884....6M}
}

@ARTICLE{Farias2019,
       author = {{Farias}, Juan P. and {Tan}, Jonathan C. and {Chatterjee}, Sourav},
        title = "{Star cluster formation from turbulent clumps. II. Gradual star cluster formation}",
      journal = {\mnras},
         year = 2019,
        month = mar,
       volume = {483},
       number = {4},
        pages = {4999-5019},
          doi = {10.1093/mnras/sty3470},
archivePrefix = {arXiv},
       eprint = {1809.04607},
 primaryClass = {astro-ph.GA},
       adsurl = {https://ui.adsabs.harvard.edu/abs/2019MNRAS.483.4999F}
}

@ARTICLE{Deacon2020,
       author = {{Deacon}, N.~R. and {Kraus}, A.~L.},
        title = "{Wide binaries are rare in open clusters}",
      journal = {\mnras},
         year = 2020,
        month = aug,
       volume = {496},
       number = {4},
        pages = {5176-5200},
          doi = {10.1093/mnras/staa1877},
archivePrefix = {arXiv},
       eprint = {2006.06679},
 primaryClass = {astro-ph.SR},
       adsurl = {https://ui.adsabs.harvard.edu/abs/2020MNRAS.496.5176D}
}

@ARTICLE{Wang2020a,
       author = {{Wang}, Long and {Nitadori}, Keigo and {Makino}, Junichiro},
        title = "{A slow-down time-transformed symplectic integrator for solving the few-body problem}",
      journal = {\mnras},
         year = 2020,
        month = apr,
       volume = {493},
       number = {3},
        pages = {3398-3411},
          doi = {10.1093/mnras/staa480},
archivePrefix = {arXiv},
       eprint = {2002.07938},
 primaryClass = {astro-ph.EP},
       adsurl = {https://ui.adsabs.harvard.edu/abs/2020MNRAS.493.3398W}
}

@ARTICLE{Wang2020b,
       author = {{Wang}, Long and {Iwasawa}, Masaki and {Nitadori}, Keigo and {Makino}, Junichiro},
        title = "{PETAR: a high-performance N-body code for modelling massive collisional stellar systems}",
      journal = {\mnras},
         year = 2020,
        month = sep,
       volume = {497},
       number = {1},
        pages = {536-555},
          doi = {10.1093/mnras/staa1915},
archivePrefix = {arXiv},
       eprint = {2006.16560},
 primaryClass = {astro-ph.IM},
       adsurl = {https://ui.adsabs.harvard.edu/abs/2020MNRAS.497..536W}
}

@ARTICLE{Schoettler2020,
       author = {{Schoettler}, Christina and {de Bruijne}, Jos and {Vaher}, Eero and {Parker}, Richard J.},
        title = "{Runaway and walkaway stars from the ONC with Gaia DR2}",
      journal = {\mnras},
         year = 2020,
        month = jul,
       volume = {495},
       number = {3},
        pages = {3104-3123},
          doi = {10.1093/mnras/staa1228},
archivePrefix = {arXiv},
       eprint = {2004.13730},
 primaryClass = {astro-ph.SR},
       adsurl = {https://ui.adsabs.harvard.edu/abs/2020MNRAS.495.3104S}
}

@ARTICLE{Hirai2021,
       author = {{Hirai}, Yutaka and {Fujii}, Michiko S. and {Saitoh}, Takayuki R.},
        title = "{SIRIUS project. I. Star formation models for star-by-star simulations of star clusters and galaxy formation}",
      journal = {\pasj},
         year = 2021,
        month = aug,
       volume = {73},
       number = {4},
        pages = {1036-1056},
          doi = {10.1093/pasj/psab038},
archivePrefix = {arXiv},
       eprint = {2005.12906},
 primaryClass = {astro-ph.GA},
       adsurl = {https://ui.adsabs.harvard.edu/abs/2021PASJ...73.1036H}
}

@ARTICLE{Fujii2021a,
       author = {{Fujii}, Michiko S. and {Saitoh}, Takayuki R. and {Wang}, Long and {Hirai}, Yutaka},
        title = "{SIRIUS project. II. A new tree-direct hybrid code for smoothed particle hydrodynamics/N-body simulations of star clusters}",
      journal = {\pasj},
         year = 2021,
        month = aug,
       volume = {73},
       number = {4},
        pages = {1057-1073},
          doi = {10.1093/pasj/psab037},
archivePrefix = {arXiv},
       eprint = {2101.05934},
 primaryClass = {astro-ph.GA},
       adsurl = {https://ui.adsabs.harvard.edu/abs/2021PASJ...73.1057F}
}

@ARTICLE{Fujii2021b,
       author = {{Fujii}, Michiko S. and {Saitoh}, Takayuki R. and {Hirai}, Yutaka and {Wang}, Long},
        title = "{SIRIUS project. III. Star-by-star simulations of star cluster formation using a direct N-body integrator with stellar feedback}",
      journal = {\pasj},
         year = 2021,
        month = aug,
       volume = {73},
       number = {4},
        pages = {1074-1099},
          doi = {10.1093/pasj/psab061},
archivePrefix = {arXiv},
       eprint = {2103.02829},
 primaryClass = {astro-ph.GA},
       adsurl = {https://ui.adsabs.harvard.edu/abs/2021PASJ...73.1074F}
}

@ARTICLE{Cournoyer-Cloutier2021,
       author = {{Cournoyer-Cloutier}, Claude and {Tran}, Aaron and {Lewis}, Sean and {Wall}, Joshua E. and {Harris}, William E. and {Mac Low}, Mordecai-Mark and {McMillan}, Stephen L.~W. and {Portegies Zwart}, Simon and {Sills}, Alison},
        title = "{Implementing primordial binaries in simulations of star cluster formation with a hybrid MHD and direct N-body method}",
      journal = {\mnras},
         year = 2021,
        month = mar,
       volume = {501},
       number = {3},
        pages = {4464-4478},
          doi = {10.1093/mnras/staa3902},
archivePrefix = {arXiv},
       eprint = {2011.06105},
 primaryClass = {astro-ph.SR},
       adsurl = {https://ui.adsabs.harvard.edu/abs/2021MNRAS.501.4464C}
}

@ARTICLE{Torniamenti2021,
       author = {{Torniamenti}, Stefano and {Ballone}, Alessandro and {Mapelli}, Michela and {Gaspari}, Nicola and {Di Carlo}, Ugo N. and {Rastello}, Sara and {Giacobbo}, Nicola and {Pasquato}, Mario},
        title = "{The impact of binaries on the evolution of star clusters from turbulent molecular clouds}",
      journal = {\mnras},
         year = 2021,
        month = oct,
       volume = {507},
       number = {2},
        pages = {2253-2266},
          doi = {10.1093/mnras/stab2238},
archivePrefix = {arXiv},
       eprint = {2104.12781},
 primaryClass = {astro-ph.GA},
       adsurl = {https://ui.adsabs.harvard.edu/abs/2021MNRAS.507.2253T}
}

@ARTICLE{Fujii2022a,
       author = {{Fujii}, Michiko S. and {Wang}, Long and {Hirai}, Yutaka and {Shimajiri}, Yoshito and {Kumamoto}, Jun and {Saitoh}, Takayuki},
        title = "{SIRIUS Project - IV. The formation history of the Orion Nebula Cluster driven by clump mergers}",
      journal = {\mnras},
         year = 2022,
        month = aug,
       volume = {514},
       number = {2},
        pages = {2513-2526},
          doi = {10.1093/mnras/stac1496},
archivePrefix = {arXiv},
       eprint = {2111.15154},
 primaryClass = {astro-ph.GA},
       adsurl = {https://ui.adsabs.harvard.edu/abs/2022MNRAS.514.2513F}
}

@ARTICLE{Fujii2022b,
       author = {{Fujii}, Michiko S. and {Hattori}, Kohei and {Wang}, Long and {Hirai}, Yutaka and {Kumamoto}, Jun and {Shimajiri}, Yoshito and {Saitoh}, Takayuki R.},
        title = "{SIRIUS Project - V. Formation of off-centre ionized bubbles associated with Orion Nebula Cluster}",
      journal = {\mnras},
         year = 2022,
        month = jul,
       volume = {514},
       number = {1},
        pages = {43-54},
          doi = {10.1093/mnras/stac808},
archivePrefix = {arXiv},
       eprint = {2206.04296},
 primaryClass = {astro-ph.GA},
       adsurl = {https://ui.adsabs.harvard.edu/abs/2022MNRAS.514...43F}
}

@INPROCEEDINGS{Offner2023,
       author = {{Offner}, S.~S.~R. and {Moe}, M. and {Kratter}, K.~M. and {Sadavoy}, S.~I. and {Jensen}, E.~L.~N. and {Tobin}, J.~J.},
        title = "{The Origin and Evolution of Multiple Star Systems}",
    booktitle = {Protostars and Planets VII},
         year = 2023,
       editor = {{Inutsuka}, S. and {Aikawa}, Y. and {Muto}, T. and {Tomida}, K. and {Tamura}, M.},
       series = {Astronomical Society of the Pacific Conference Series},
       volume = {534},
        month = jul,
        pages = {275},
          doi = {10.48550/arXiv.2203.10066},
archivePrefix = {arXiv},
       eprint = {2203.10066},
 primaryClass = {astro-ph.SR},
       adsurl = {https://ui.adsabs.harvard.edu/abs/2023ASPC..534..275O}
}

@ARTICLE{Cournoyer-Cloutier2023,
       author = {{Cournoyer-Cloutier}, Claude and {Sills}, Alison and {Harris}, William E. and {Appel}, Sabrina M. and {Lewis}, Sean C. and {Polak}, Brooke and {Tran}, Aaron and {Wilhelm}, Maite J.~C. and {Mac Low}, Mordecai-Mark and {McMillan}, Stephen L.~W. and et al.},
        title = "{Early evolution and three-dimensional structure of embedded star clusters}",
      journal = {\mnras},
         year = 2023,
        month = may,
       volume = {521},
       number = {1},
        pages = {1338-1352},
          doi = {10.1093/mnras/stad568},
archivePrefix = {arXiv},
       eprint = {2302.08536},
 primaryClass = {astro-ph.GA},
       adsurl = {https://ui.adsabs.harvard.edu/abs/2023MNRAS.521.1338C}
}

@ARTICLE{Cournoyer-Cloutier2024_ClusterAssembly,
       author = {{Cournoyer-Cloutier}, Claude and {Karam}, Jeremy and {Sills}, Alison and {Portegies Zwart}, Simon and {Wilhelm}, Maite J.~C.},
        title = "{Binary Disruption and Ejected Stars from Hierarchical Star Cluster Assembly}",
      journal = {\apj},
         year = 2024,
        month = nov,
       volume = {975},
       number = {2},
          eid = {207},
        pages = {207},
          doi = {10.3847/1538-4357/ad7f50},
archivePrefix = {arXiv},
       eprint = {2409.13564},
 primaryClass = {astro-ph.GA},
       adsurl = {https://ui.adsabs.harvard.edu/abs/2024ApJ...975..207C}
}

@ARTICLE{Cournoyer-Cloutier2024,
       author = {{Cournoyer-Cloutier}, Claude and {Sills}, Alison and {Harris}, William E. and {Polak}, Brooke and {Rieder}, Steven and {Andersson}, Eric P. and {Appel}, Sabrina M. and {Mac Low}, Mordecai-Mark and {McMillan}, Stephen and {Portegies Zwart}, Simon},
        title = "{Massive Star Cluster Formation with Binaries. I. Evolution of Binary Populations}",
      journal = {\apj},
         year = 2024,
        month = dec,
       volume = {977},
       number = {2},
          eid = {203},
        pages = {203},
          doi = {10.3847/1538-4357/ad90b3},
archivePrefix = {arXiv},
       eprint = {2410.07433},
 primaryClass = {astro-ph.GA},
       adsurl = {https://ui.adsabs.harvard.edu/abs/2024ApJ...977..203C}
}

@ARTICLE{Rantala2024,
       author = {{Rantala}, Antti and {Naab}, Thorsten and {Lah{\'e}n}, Natalia},
        title = "{FROST-CLUSTERS - I. Hierarchical star cluster assembly boosts intermediate-mass black hole formation}",
      journal = {\mnras},
         year = 2024,
        month = jul,
       volume = {531},
       number = {3},
        pages = {3770-3799},
          doi = {10.1093/mnras/stae1413},
archivePrefix = {arXiv},
       eprint = {2403.10602},
 primaryClass = {astro-ph.GA},
       adsurl = {https://ui.adsabs.harvard.edu/abs/2024MNRAS.531.3770R}
}

@ARTICLE{Karam2025,
       author = {{Karam}, Jeremy and {Fujii}, Michiko S. and {Sills}, Alison},
        title = "{Dynamics of Star Cluster Formation: The Effects of Ongoing Star Formation and Stellar Feedback}",
      journal = {\apj},
         year = 2025,
        month = may,
       volume = {984},
       number = {1},
          eid = {75},
        pages = {75},
          doi = {10.3847/1538-4357/adc719},
archivePrefix = {arXiv},
       eprint = {2503.21716},
 primaryClass = {astro-ph.GA},
       adsurl = {https://ui.adsabs.harvard.edu/abs/2025ApJ...984...75K}
}

@ARTICLE{Amiri2026,
       author = {{Amiri}, Vahid and {Flammini Dotti}, Francesco and {Pang}, Xiaoying and {Kamlah}, A.~W.~H. and {Berczik}, Peter and {Shukirgaliyev}, Bekdaulet and {Spurzem}, Rainer},
        title = "{Primordial Binary Stars, Mass segregation and Fractality Effects on the Early Evolution of Young Open Clusters}",
      journal = {arXiv e-prints},
         year = 2026,
        month = jun,
          eid = {arXiv:2606.04509},
        pages = {arXiv:2606.04509},
          doi = {10.48550/arXiv.2606.04509},
archivePrefix = {arXiv},
       eprint = {2606.04509},
 primaryClass = {astro-ph.GA},
       adsurl = {https://ui.adsabs.harvard.edu/abs/2026arXiv260604509A}
}

@software{portegies_zwart_2023_8409512,
  author       = {Portegies Zwart, Simon and
                  van Elteren, Arjen and
                  Pelupessy, Inti and
                  McMillan, Steve and
                  Rieder, Steven and
                  de Vries, Nathan and
                  Marosvolgyi, Marcell and
                  Whitehead, Alfred and
                  Wall, Joshua and
                  Drost, Niels and
                  Jilkova, Lucie and
                  Martinez Barbosa, Carmen and
                  van der Helm, Edwin and
                  Beedorf, Jeroen and
                  Bos, Patrick and
                  Boekholt, Tjarda and
                  van Werkhoven, Ben and
                  Wijnen, Thomas and
                  Hamers, Adrian and
                  Caputo, Daniel and
                  Ferrari, Guilherme and
                  Toonen, Silvia and
                  Gaburov, Evghenii and
                  Paardekooper, Jan-Pieter and
                  Janes, Jurgen and
                  Punzo, Davide and
                  Kruip, Chael and
                  Altay, Gabriel},
  title        = {AMUSE: the Astrophysical Multipurpose Software
                   Environment
                  },
  month        = oct,
  year         = 2023,
  publisher    = {Zenodo},
  version      = {v2023.7.0},
  doi          = {10.5281/zenodo.8409512},
  url          = {https://doi.org/10.5281/zenodo.8409512},
}

@BOOK{Portegies2018,
       author = {{Portegies Zwart}, Simon and {McMillan}, Steve},
        title = "{Astrophysical Recipes; The art of AMUSE}",
         year = 2018,
          doi = {10.1088/978-0-7503-1320-9},
       adsurl = {https://ui.adsabs.harvard.edu/abs/2018araa.book.....P}
}

@ARTICLE{Pelupessy2013,
       author = {{Pelupessy}, F.~I. and {van Elteren}, A. and {de Vries}, N. and {McMillan}, S.~L.~W. and {Drost}, N. and {Portegies Zwart}, S.~F.},
        title = "{The Astrophysical Multipurpose Software Environment}",
      journal = {\aap},
         year = 2013,
        month = sep,
       volume = {557},
          eid = {A84},
        pages = {A84},
          doi = {10.1051/0004-6361/201321252},
archivePrefix = {arXiv},
       eprint = {1307.3016},
 primaryClass = {astro-ph.IM},
       adsurl = {https://ui.adsabs.harvard.edu/abs/2013A&A...557A..84P}
}

@ARTICLE{Portegies2013,
       author = {{Portegies Zwart}, S. and {McMillan}, S.~L.~W. and {van Elteren}, E. and {Pelupessy}, I. and {de Vries}, N.},
        title = "{Multi-physics simulations using a hierarchical interchangeable software interface}",
      journal = {Computer Physics Communications},
         year = 2013,
        month = mar,
       volume = {184},
       number = {3},
        pages = {456-468},
          doi = {10.1016/j.cpc.2012.09.024},
archivePrefix = {arXiv},
       eprint = {1204.5522},
 primaryClass = {astro-ph.IM},
       adsurl = {https://ui.adsabs.harvard.edu/abs/2013CoPhC.184..456P}
}

@ARTICLE{Portegies2009,
       author = {{Portegies Zwart}, Simon and {McMillan}, Steve and {Harfst}, Stefan and {Groen}, Derek and {Fujii}, Michiko and {Nuall{\'a}in}, Breannd{\'a}n {\'O}. and {Glebbeek}, Evert and {Heggie}, Douglas and {Lombardi}, James and {Hut}, Piet and et al.},
        title = "{A multiphysics and multiscale software environment for modeling astrophysical systems}",
      journal = {\na},
         year = 2009,
        month = may,
       volume = {14},
       number = {4},
        pages = {369-378},
          doi = {10.1016/j.newast.2008.10.006},
archivePrefix = {arXiv},
       eprint = {0807.1996},
 primaryClass = {astro-ph},
       adsurl = {https://ui.adsabs.harvard.edu/abs/2009NewA...14..369P}
}

@ARTICLE{Hurley2000,
       author = {{Hurley}, Jarrod R. and {Pols}, Onno R. and {Tout}, Christopher A.},
        title = "{Comprehensive analytic formulae for stellar evolution as a function of mass and metallicity}",
      journal = {\mnras},
         year = 2000,
        month = jul,
       volume = {315},
       number = {3},
        pages = {543-569},
          doi = {10.1046/j.1365-8711.2000.03426.x},
archivePrefix = {arXiv},
       eprint = {astro-ph/0001295},
 primaryClass = {astro-ph},
       adsurl = {https://ui.adsabs.harvard.edu/abs/2000MNRAS.315..543H}
}

@ARTICLE{Hurley2002,
       author = {{Hurley}, Jarrod R. and {Tout}, Christopher A. and {Pols}, Onno R.},
        title = "{Evolution of binary stars and the effect of tides on binary populations}",
      journal = {\mnras},
         year = 2002,
        month = feb,
       volume = {329},
       number = {4},
        pages = {897-928},
          doi = {10.1046/j.1365-8711.2002.05038.x},
archivePrefix = {arXiv},
       eprint = {astro-ph/0201220},
 primaryClass = {astro-ph},
       adsurl = {https://ui.adsabs.harvard.edu/abs/2002MNRAS.329..897H}
}

@ARTICLE{Tanikawa2020,
       author = {{Tanikawa}, Ataru and {Yoshida}, Takashi and {Kinugawa}, Tomoya and {Takahashi}, Koh and {Umeda}, Hideyuki},
        title = "{Fitting formulae for evolution tracks of massive stars under extreme metal-poor environments for population synthesis calculations and star cluster simulations}",
      journal = {\mnras},
         year = 2020,
        month = jul,
       volume = {495},
       number = {4},
        pages = {4170-4191},
          doi = {10.1093/mnras/staa1417},
archivePrefix = {arXiv},
       eprint = {1906.06641},
 primaryClass = {astro-ph.HE},
       adsurl = {https://ui.adsabs.harvard.edu/abs/2020MNRAS.495.4170T}
}

@misc{pynbody,
  author = {{Pontzen}, A. and {Ro{\v s}kar}, R. and {Stinson}, G.~S. and {Woods},
     R. and {Reed}, D.~M. and {Coles}, J. and {Quinn}, T.~R.},
  title = "{pynbody: Astrophysics Simulation Analysis for Python}",
  note = {Astrophysics Source Code Library, ascl:1305.002},
  year = 2013
}

\begin{appendix}
\end{appendix}

\end{document}